\documentclass[useAMS,usenatbib]{mnras}

\usepackage{times}
\usepackage{graphicx}
\usepackage{subfigure}
\usepackage{amsmath}

\newcommand{\gsim}{\mathrel{\hbox{\rlap{\lower.55ex \hbox {$\sim$}}
                   \kern-.3em \raise.4ex \hbox{$>$}}}}
\newcommand{\lsim}{\mathrel{\hbox{\rlap{\lower.55ex \hbox {$\sim$}}
                   \kern-.3em \raise.4ex \hbox{$<$}}}}

\title[Stellar properties at $z=5$ ]{The statistical properties of stars at redshift, $z=5$, compared with the present epoch}
\author[M. R. Bate]{Matthew R. Bate$^{1}$\thanks{E-mail:
M.R.Bate@exeter.ac.uk}\\
$^{1}$ Department of Physics and Astronomy, University of Exeter, Stocker
Road, Exeter EX4 4QL, United Kingdom
}

\date{Accepted by MNRAS}
\begin{document}
\maketitle
\begin{abstract}
We report the statistical properties of stars and brown dwarfs obtained from three radiation hydrodynamical simulations of star cluster formation with metallicities of 1, 1/10 and 1/100 of the solar value.  The star-forming clouds are subjected to cosmic microwave background radiation that is appropriate for star formation at a redshift $z=5$. The results from the three calculations are compared to each other, and to similar previously published calculations that had levels of background radiation appropriate for present-day ($z=0$) star formation.  Each of the calculations treat dust and gas temperatures separately and include a thermochemical model of the diffuse interstellar medium.  We find that whereas the stellar mass distribution is insensitive to the metallicity for present-day star formation, at $z=5$ the characteristic stellar mass increases with increasing metallicity and the mass distribution has a deficit of brown dwarfs and low-mass stars at solar metallicity compared to the Galactic initial mass function.  We also find that the multiplicity of M-dwarfs decreases with increasing metallicity at $z=5$.  These effects are a result of metal-rich gas being unable to cool to as low temperatures at $z=5$ compared to at $z=0$ due to the hotter cosmic microwave background radiation, which inhibits fragmentation at high densities.

\end{abstract}

\section{Introduction}
\label{introduction}

Around a decade ago, numerical studies of present day star formation advanced to the point that individual radiation hydrodynamical calculations could produce stellar groups or clusters containing more than a hundred stars and brown dwarfs.  With such numbers of stars and brown dwarfs, the statistical properties of the stellar populations can be meaningfully compared with those of observed Galactic stellar populations.  \cite{Bate2012} published the first radiation hydrodynamical calculation to reproduce a wide variety of observed statistical properties of Galactic stellar systems.  The calculation produced in excess of 180 stars and brown dwarfs, including 40 multiple systems, while also resolving young discs down to radii of $\approx 10$~au.  The stellar mass function was found to be in good agreement with the observed Galactic initial mass function (IMF), the stellar multiplicity as a function of primary mass was in agreement with the results from field star surveys, and many of the observed characteristics of multiple stellar systems were reproduced.  Moreover, \cite{Bate2018} analysed the mass and radius distributions of the young discs that were produced by the \cite{Bate2012} calculation and found that the distribution of disc radii are in good agreement with the observed radii of discs in Galactic star-forming regions, and the mass distributions are similar to the estimated mass distributions of very young (Class 0) protostellar systems \citep[see][]{Tychoniec_etal2018,Tobin_etal2020}.

In addition to comparing the results of star cluster formation calculations to observed Galactic stellar populations, the ability to perform numerical simulations that produce clusters of stars gives us the ability to directly predict how the statistical properties of stellar systems depend on environment and initial conditions, and to investigate the roles of various physical processes.  
Thus far, hydrodynamical calculations that either include radiative transfer and/or radiative heating from protostars have been used to explore the dependence of stellar properties on metallicity (\citealt{Myersetal2011, Bate2014, Bate2019}; \citealt*{HeRicGee2019}; \citealt{Guszejnov_etal2022}), molecular cloud density \citep{Bate2009b, JonBat2018a,TanKruFed2022,Guszejnov_etal2022}, protostellar outflows (\citealt*{KruKleMcK2012, MatFed2021,TanKruFed2022}; \citealt{Grudic_etal2021a,Grudic_etal2022,Guszejnov_etal2021,Guszejnov_etal2022}), variations in turbulence (\citealt*{LomWhiHub2015,NamFedKru2021,MatFedSet2022}; \citealt{Guszejnov_etal2022}), and magnetic fields (\citealt{Myersetal2013, Myers_etal2014, Krumholz_etal2016, Cunningham_etal2018,Grudic_etal2021a,Guszejnov_etal2022}).  
Some of the most recent calculations have even begun exploring how the non-ideal magnetohydrodynamic (MHD) effects of ambipolar diffusion, Hall conduction, and Ohmic resistivity affect the formation of stellar groups, albeit only for small numbers of stars as yet \citep{WurBatPri2019}.

The study of the dependence of star formation on metallicity of \citet{Bate2019} employed the method of \citet{BatKet2015} to model the thermodynamics.  This method combines radiative transfer with a thermochemical model of the diffuse interstellar medium (ISM), with gas and dust temperatures treated separately.  The thermochemical model includes simple chemical models for the atomic and molecular forms of hydrogen and whether carbon is in the form of C$^+$, neutral carbon, or CO.  The ISM model is similar to that used by \citet{GloCla2012c} to study molecular cloud evolution, although the chemical model employed by \citeauthor{BatKet2015} is greatly simplified.  \citet{Bate2019} found that the stellar properties did not vary greatly for present-day star formation if the metallicities ranged from 1/100 to 3 times the solar value.  For example, the stellar mass functions were statistically indistinguishable.  However, the greater cooling rates at high gas densities due to the lower opacities at low metallicities were found to increase the fragmentation on small spatial scales which resulted in an anti-correlation between the close binary fraction of low-mass stars and metallicity similar to that which is observed (\citealt{Badenes_etal2018}; \citealt*{ElBRix2019,MoeKraBad2019}).  The fraction of protostellar mergers at low metallicities also increased due to the higher prevalence of small-scale fragmentation.  An indication was found that at lower metallicity close binaries may have lower mass ratios and the relative abundance of brown dwarfs to stars may increase slightly, but these effects were quite weak.  \cite{ElsBat2021} also studied the discs produced by the calculations and found that the characteristic radii of protostellar discs decrease with decreasing metallicity and that the discs and orbits of pairs of protostars tend to be less well aligned at lower metallicity.

In this paper, we report results from three new radiation hydrodynamical calculations of stellar cluster formation that are identical to those reported by \cite{Bate2019} except that the background interstellar radiation field (ISRF) is modified so as to include a cosmic microwave background radiation component that is appropriate for a redshift $z=5$ rather than $z=0$.  We follow the collapse of each of the molecular clouds to form a cluster of stars and then compare the properties of the stars and brown dwarfs, both between each of the new calculations, and to the results from the present-day ($z=0$) calculations of \citet{Bate2019}.  In Section \ref{sec:method}, we summarise the method and initial conditions. The results from the calculations are presented in Section \ref{sec:results}.  In Section \ref{sec:discussion}, we discuss observational evidence for variation of the stellar mass function in the context of the results presented in this paper.  Section \ref{conclusions} contains the conclusions.

\section{Method}
\label{sec:method}

The calculations were performed using the smoothed particle
hydrodynamics (SPH) code, {\sc sphNG}, based on the original 
version from \citeauthor{Benz1990} 
(\citeyear{Benz1990}; \citealt{Benzetal1990}), but substantially
modified using the methods described in \citet{BatBonPri1995}, \citet{PriMon2007},
\citet*{WhiBatMon2005}, \citet{WhiBat2006}, \cite{BatKet2015} and 
parallelised using both MPI and OpenMP.

Gravitational forces and particle nearest neighbours were calculated using a binary tree.  
The smoothing lengths of particles were set such that the smoothing
length of each particle 
$h = 1.2 (m/\rho)^{1/3}$ where $m$ and $\rho$ are the 
SPH particle's mass and density, respectively
\cite[see][for further details]{PriMon2007}.  A second-order Runge-Kutta-Fehlberg 
integrator \citep{Fehlberg1969} was employed, with individual time steps for each particle
\citep*{BatBonPri1995}.
The \cite{MorMon1997} artificial viscosity was used to reduce numerical shear viscosity, 
with $\alpha_{\rm_v}$ varying between 0.1 and 1 and setting $\beta_{\rm v}=2 \alpha_{\rm v}$
\citep[see also][]{PriMon2005}.

\subsection{Radiative transfer and the diffuse ISM model}
\label{hydro}

The calculations used the combined radiative transfer and diffuse ISM thermochemical model 
developed by \cite{BatKet2015}.  A brief summary of the main elements of
the method is provided by \cite{Bate2019}; the reader is directed to those two papers
for the details of the method since the calculations performed for the 
present paper are identical to the calculations presented by \cite{Bate2019} except for the
form of the interstellar radiation field.  

Briefly, the radiation hydrodynamical calculations are solved in a frame co-moving with the fluid, assuming local thermal equilibrium
(LTE), and can be written
\begin{equation}
\label{rhd1}
\frac{{\rm D}\rho}{{\rm D}t} + \rho \nabla\cdot \mbox{\boldmath $v$} = 0~,
\end{equation}
\begin{equation}
\label{rhd2}
\rho \frac{{\rm D}\mbox{\boldmath $v$}}{{\rm D}t} = -\nabla p + \frac{\mbox{${\chi}\rho$}}{c} \mbox{\boldmath $F$}~,
\end{equation}
\begin{equation}
\label{rhdnew3}
\begin{split}
\rho \frac{{\rm D}}{{\rm D}t}\left( \frac{E}{\rho}\right) = - & \nabla\cdot \mbox{\boldmath $F$} - \nabla\mbox{\boldmath $ v${\bf :P}} - a c \kappa_{\rm g} \rho \left(T_{\rm r}^4 - T_{\rm g}^4 \right) - \\ & a c \kappa_{\rm d} \rho  \left(T_{\rm r}^4 - T_{\rm d}^4 \right),
\end{split}
\end{equation}
\begin{equation}
\label{rhdnew4}
\begin{split}
\rho \frac{{\rm D}u}{{\rm D}t} = - & p \nabla\cdot \mbox{\boldmath $v$} + a c \kappa_{\rm g} \rho  \left(T_{\rm r}^4 - T_{\rm g}^4 \right) - \\ &   \Lambda_{\rm gd} - \Lambda_{\rm line} + \Gamma_{\rm cr} + \Gamma_{\rm pe} + \Gamma_{\rm H2,g}~,
\end{split}
\end{equation}
\citep{BatKet2015}, where, ${\rm D}/{\rm D}t \equiv \partial/\partial t + \mbox{\boldmath $v$}\cdot \nabla$ is the convective derivative.  The symbols $\rho, $ \mbox{\boldmath $v$} and $p$ represent the material mass density, velocity, and scalar isotropic pressure, respectively, and $c$ is the speed of light.  The total frequency-integrated radiation density, momentum density (flux) and pressure tensor are represented by $E$, {\boldmath $F$}, and {\bf P}, respectively.  The flux-limited diffusion approximation is employed, such that $\mbox{\boldmath $F$} = -c\lambda \nabla E/(\chi \rho)$ and ${\bf P}={\bf f}E$, in which the flux limiter, $\lambda$, of \citet{LevPom1981} is used and the Eddington tensor, ${\bf f}$, is computed using the gradient of $E$. Equations \ref{rhdnew3}--\ref{rhdnew4} are used to evolve the continuum radiation field and the specific internal energy of the gas, $u$, separately from the dust.  The gas, dust, and continuum radiation fields all have their own associated temperatures: $T_{\rm g}$, $T_{\rm d}$, and $T_{\rm r}$, respectively.  The dust temperature is set by assuming that the dust is in thermal equilibrium with the combined ISRF and continuum radiation field, $E$, taking into account the transfer of energy between the gas and the dust:
\begin{equation}
\rho \Lambda_{\rm ISR} + ac\kappa_{\rm d}\rho(T_{\rm r}^4 - T_{\rm d}^4) + \Lambda_{\rm gd} = 0,
\end{equation}
where $\Lambda_{\rm ISR}$ is the heating rate per unit mass of the dust due to the ISRF.
The terms at the end of equation \ref{rhdnew4} are  $\Lambda_{\rm gd}$ which accounts for the thermal interaction due to collisions between the gas and the dust, $\Lambda_{\rm line}$ which is the cooling rate per unit volume due to atomic and molecular line emission, $\Gamma_{\rm cr}$ which is the heating rate per unit volume due to cosmic rays,  $\Gamma_{\rm pe}$ which is the heating rate per unit volume due to the photoelectric effect, and $\Gamma_{\rm H2,g}$ which is the heating rate per unit volume due to the formation of molecular hydrogen on dust grains. We also define the radiation constant $a=4 \sigma_{\rm B}/c$, where $\sigma_{\rm B}$ is the Stefan-Boltzmann constant.  The Rosseland mean gas opacity $\kappa_{\rm g}$ is only important above the dust sublimation temperature, and the mean dust opacity is $\kappa_{\rm d}$.   The total opacity, $\chi$, in equation \ref{rhd2} and the equation for $\mbox{\boldmath $F$}$ is set to the sum of the gas and dust opacities and ignores scattering.  It may be noted that the radiation pressure term (the last term in equation \ref{rhd2}) and the second term on the right-hand side of equation \ref{rhdnew3} are both negligible in all of the calculations discussed in this paper.

\begin{figure}
\vspace{-0.5cm}
\centering
    \includegraphics[width=9cm]{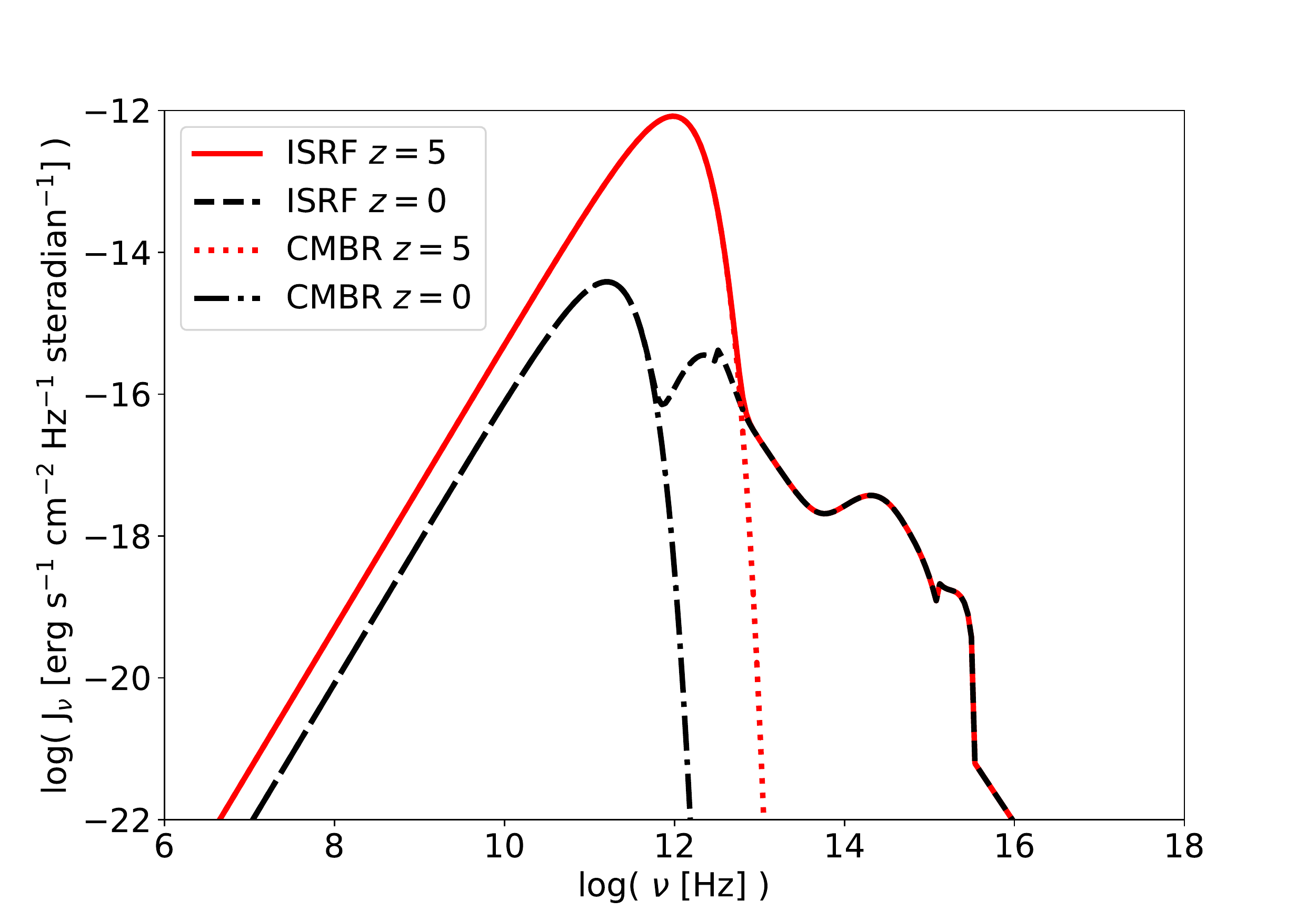} \vspace{-0.5cm}
\caption{A comparison of the form of the interstellar radiation field (ISRF) that is assumed at redshift $z=5$ compared to that at $z=0$.  It is assumed that the only difference is due to the cosmic microwave background radiation (CMBR) which has a greater temperature at $z=5$ (the contributions of the CMBR to each ISRF are shown separately).}
\label{ISRF}
\end{figure}

The hydrogen and helium mass fractions were taken to be $X=0.70$ and $Y=0.28$, respectively, 
with the solar metallicity abundance set to be ${\rm Z}_\odot=0.02$.
We employed simple chemical models to treat the evolution of hydrogen and carbon.
The abundances of C$+$, neutral carbon, CO, and the depletion of CO on to dust grains
were computed using the model of \cite{KetCas2008}.  For hydrogen, we evolved the
atomic and molecular hydrogen abundances using the same molecular hydrogen 
formation and dissociation rates as those used by \cite{Gloveretal2010}.

The clouds were assumed to be bathed in an ISRF.  
This contributes to both the heating rate of dust grains and photoelectric heating of the gas.  
The ISRF is attenuated due to dust extinction inside a molecular cloud, with opacities set in the same way as in \cite{Bate2014, Bate2019}.
To describe the ISRF at the present epoch (redshift $z=0$) the analytic form of \cite*{ZucWalGal2001} is used, but modified to
include the `standard' UV component from \cite{Draine1978} in the energy range $h\nu=5-13.6$~eV.
For the calculations presented in this paper, we modified the functional form of the component
of the ISRF that is due to the cosmic microwave background radiation.  Specifically, we take the temperature of the cosmic microwave background radiation (CMBR) to scale as
\begin{equation}
T_{\rm CMBR}(z) = (1+z) T_{\rm CMBR}(0),
\label{eq:TCMBR}
\end{equation}
where $T_{\rm CMBR}(0)=2.73$~K, so that at $z=5$, $T_{\rm CMBR}(z) = 16.4$~K.
In Figure \ref{ISRF}, we plot the forms of the ISRF that \cite{Bate2019} used for the present day (redshift $z=0$) and the form that we use for redshift $z=5$.  The contributions of the CMBR to each of the ISRFs are also plotted in the figure.  Note that we assume that the other contributions to the ISRF do not change with redshift.  Essentially we are assuming that the star-forming clouds are in similar radiative environments, except for the contribution from the CMBR.  This need not be the case -- for example, if a star-forming cloud was in a galactic starburst environment there may also be a stronger high-frequency component to the ISRF.  However, making such additional changes to the ISRF would add more free parameters to the models.  Therefore, for the present paper we only consider the effect of the hotter/stronger CMBR.

\subsection{Sink particles}
\label{sec:sinks}

As in \cite{Bate2019}, the calculations followed the hydrodynamic collapse of each protostar 
through the first core phase and into the second collapse phase (that begins at densities of
$\sim 10^{-7}$~g~cm$^{-3}$) due to molecular hydrogen dissociation \citep{Larson1969}.
Sink particles \citep{BatBonPri1995} were inserted when the density exceeded
$10^{-5}$~g~cm$^{-3}$.  This density is 
only a factor of one hundred lower than the density at which
stellar core begins to form (density $\sim 10^{-3}$~g~cm$^{-3}$) and the free-fall time
at this density is only one week.

The sink particle accretion radius was set to $r_{\rm acc}=0.5$ au.
Sink particles interact with the gas only via gravity and accretion, and the gravitational interaction between sink particles is not softened.
Sink particle trajectories are integrated using the same Runge-Kutta-Fehlberg 
integrator that is used for the SPH particles, but with a much lower error 
tolerance so that even the closest orbits (semi-major axes less than 1 au) are maintained for the period of the simulations (i.e. $\sim 10^5$~yrs).
The sink particles themselves do not contribute radiative feedback 
\citep[see][for discussion of this limitation]{Bate2012,JonBat2018b}.
Sink particles are merged together if they
pass within 0.03 au (i.e., $\approx 6$~R$_\odot$).

\begin{table*}
\begin{tabular}{lcccccccccccc}\hline
Redshift  & Initial Gas & Metallicity & No. Stars & No. Brown  & Mass of Stars \&  & Mean  & Mean & Median & Stellar\\
& Mass  &   & Formed & Dwarfs  & Brown Dwarfs & Mass & Log-Mass & Mass & Mergers \\
&  M$_\odot$ &  Z$_\odot$ & & & M$_\odot$ & M$_\odot$ &M$_\odot$ & M$_\odot$ \\ 
\hline
$z=5$ & 500 &  0.01 & $\geq 89$& $\leq 43$& 52.4 & $0.40\pm0.05$ & $0.16\pm0.02$ & 0.15 & 17 \\  
& 500 &  0.1 & $\geq 81 $& $\leq 33$& 56.5 & $0.50\pm0.07$ & $0.20\pm0.03$ & 0.19 & 7  \\   
& 500 &  1.0 & $\geq 65 $& $\leq 2$& 67.4 & $1.01\pm0.17$ & $0.49\pm0.08$ & 0.40 & 1 \\  
\hline
$z=0$ & 500 &  0.01 & $\geq 90$& $\leq 52$& 49.8 & $0.35\pm0.04$ & $0.15\pm0.02$ & 0.14 & 17 \\   
& 500 &  0.1 & $\geq 125$& $\leq 49$& 73.4 & $0.42\pm0.05$ & $0.17\pm0.02$ & 0.15 & 11 \\   
& 500 &  1.0 & $\geq 171$& $\leq 84$& 90.1 & $0.35\pm0.04$ & $0.16\pm0.01$ & 0.15 & 14\\  
& 500 & 3.0 & $\geq 193$& $\leq 65$& 92.0 & $0.36\pm0.03$ & $0.18\pm0.01$ & 0.17 & 6 \\   
\hline
\end{tabular}
\caption{\label{table1} The parameters and overall statistical results for each of the three new radiation hydrodynamical calculations performed for this paper, and the four calculations from \citet{Bate2019}.  All calculations were run to 1.20~$t_{\rm ff}$.  Brown dwarfs are defined as having final masses less than 0.075 M$_\odot$ [Note that the equivalent table in \citet{Bate2019} mistakenly gave incorrect numbers of stars and brown dwarfs, although the sums of these numbers (the total numbers of objects) were correct.].  The numbers of stars (brown dwarfs) are lower (upper) limits because some brown dwarfs were still accreting when the calculations were stopped.  At $z=0$, different metallicities result in no significant difference in the mean and median masses of the stellar populations, but with $z=5$ the solar-metallicity calculation produces a much heavier population of stars.  At both redshifts, less gas is converted into stars in calculations with lower metallicities and the average number of stellar mergers per star consistently increases with decreasing metallicity.  This table and caption are comparable to Table 1 in \citet{Bate2019}.}
\end{table*}

\begin{figure*}
\centering \vspace{-0.25cm}
    \includegraphics[width=16cm]{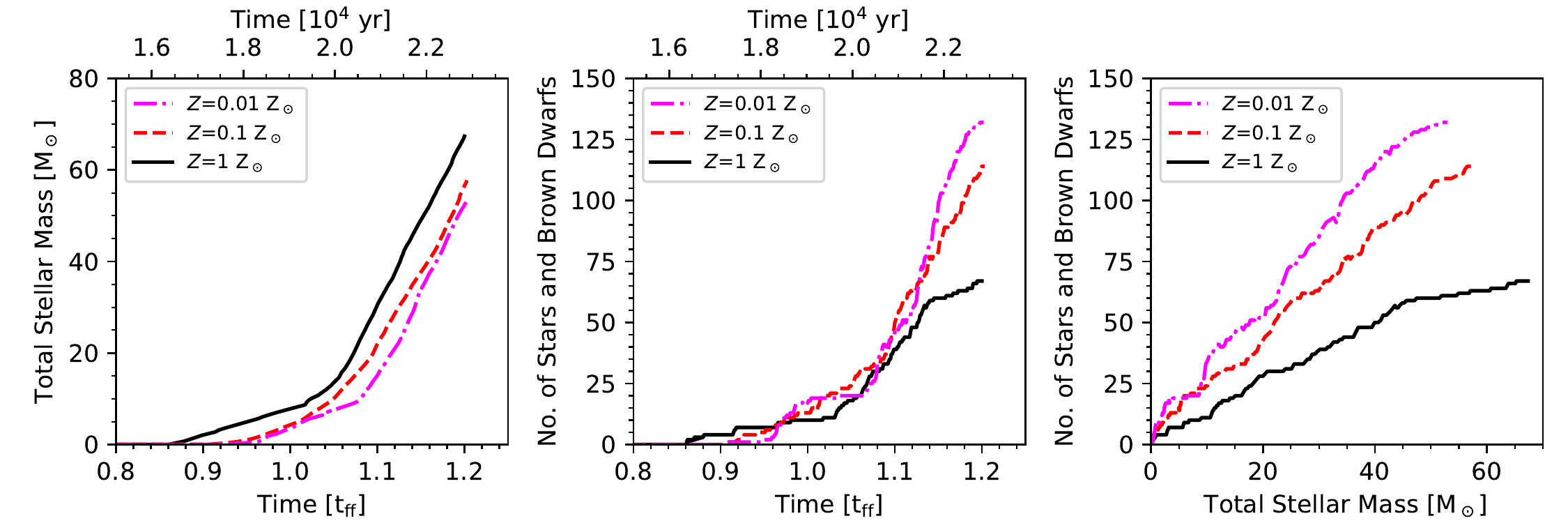} \vspace{-0.1cm}
\caption{The star formation rates obtained for each of the three radiation hydrodynamical calculations.  We plot: the total stellar mass (i.e., the mass contained in sink particles) versus time (left panel), the number of stars and brown dwarfs (i.e., the number of sink particles) versus time (centre panel), and the number of stars and brown dwarfs versus the total stellar mass (right panel).  The different line types are for metallicities of 1/100 (magenta, dot-dashed), 1/10 (red, dashed), and 1 (black, solid) times solar. Time is measured from the beginning of the calculation in terms of the free-fall time of the initial cloud (bottom) or years (top).  The star formation is delayed in the two low-metallicity calculations, due to the warmer gas temperatures, but after $t\approx 1.08$~$t_{\rm ff}$ the star formation rates in terms of the amount of gas converted to stars are very similar for all of the calculations.  There is also a clear progression such that more objects are produced from a given amount of gas for lower metallicities (right panel). This figure and caption are comparable to Fig.~1, \citet{Bate2019}.  }
\label{massnumber}
\end{figure*}

\subsection{Initial conditions and resolution}
\label{initialconditions}

The initial density and velocity structure for each calculation were identical to those used by \cite{Bate2012, Bate2014, Bate2019} -- this allows close comparison of the results with those of \citet{Bate2019} because only one or two parameters are changed at a time (i.e.\ the metallicity and/or the redshift).  We only provide a brief description of the initial conditions here -- see \cite{Bate2012} for a more complete description.  In short, for each calculation the initial conditions consisted of a uniform-density, spherical cloud containing 
500 M$_\odot$ of molecular gas, with a radius of 0.404 pc (83300 au).  This gives an initial density of 
$1.2\times 10^{-19}$~g~cm$^{-3}$ ($n_{\rm H} \approx 6\times 10^4$~cm$^{-3}$) and an initial free-fall time of 
$t_{\rm ff}=6.0\times 10^{12}$~s or $1.90\times 10^5$ years.    
Although each cloud was uniform in density, we imposed an initial 
supersonic `turbulent' velocity field in the same manner
as \citet*{OstStoGam2001} and \cite*{BatBonBro2003}.
The power spectrum of the divergence-free random Gaussian velocity field was $P(k) \propto k^{-4}$, where $k$ is the wavenumber. 
The velocities were set on a $128^3$ uniform grid and the SPH particle velocities were interpolated from the grid.
The velocity field was normalised so that the kinetic energy 
of the turbulence was equal to the magnitude of the gravitational potential 
energy of the cloud (this results in the initial root-mean-square Mach number of the turbulence being ${\cal M}=11.2$ at 15~K).  The same velocity field was used for each calculation.

As in \cite{Bate2019}, the initial gas and dust temperatures were set such that the dust was in thermal equilibrium with the local interstellar radiation field.  The gas was in thermal equilibrium such that heating from the ISRF and cosmic rays was balanced by cooling from atomic and molecular line emission and collisional coupling with the dust.  This produces clouds with dust temperatures that are warmest on the outside and coldest at the centre.  For the solar metallicity cloud, the initial dust temperature ranges from $T_{\rm dust}=16.3-17.9$~K. For $Z=0.1~{\rm Z}_\odot$, $T_{\rm dust}=16.8-19.1$~K and for the lowest metallicity, $T_{\rm dust}=18.6-19.9$~K. The mean initial gas temperatures for the three calculations were $T_{\rm gas} \approx 16.5, 17.6, 14.3$~K for metallicities $Z=1,0.1, 0.01~{\rm Z}_\odot$, respectively.

The calculations used $3.5 \times 10^7$ SPH particles to model the cloud (the same resolution as that used by \citealt{Bate2019}).  This resolution is sufficient to resolve the local Jeans mass throughout the calculation, which is necessary to model fragmentation of collapsing molecular clouds correctly (\citealt{BatBur1997, Trueloveetal1997, Whitworth1998, Bossetal2000}; \citealt*{HubGooWhi2006}).  Currently there is no consensus as to the resolution that is necessary and sufficient to capture disc fragmentation (see Section 2.3 of \citealt{Bate2019} for discussion of this point and references for further reading).  The fact that the criteria for disc fragmentation are not well understood should be kept in mind as a caveat for the results presented below.

\begin{figure*}
\centering
    \includegraphics[width=17.5cm]{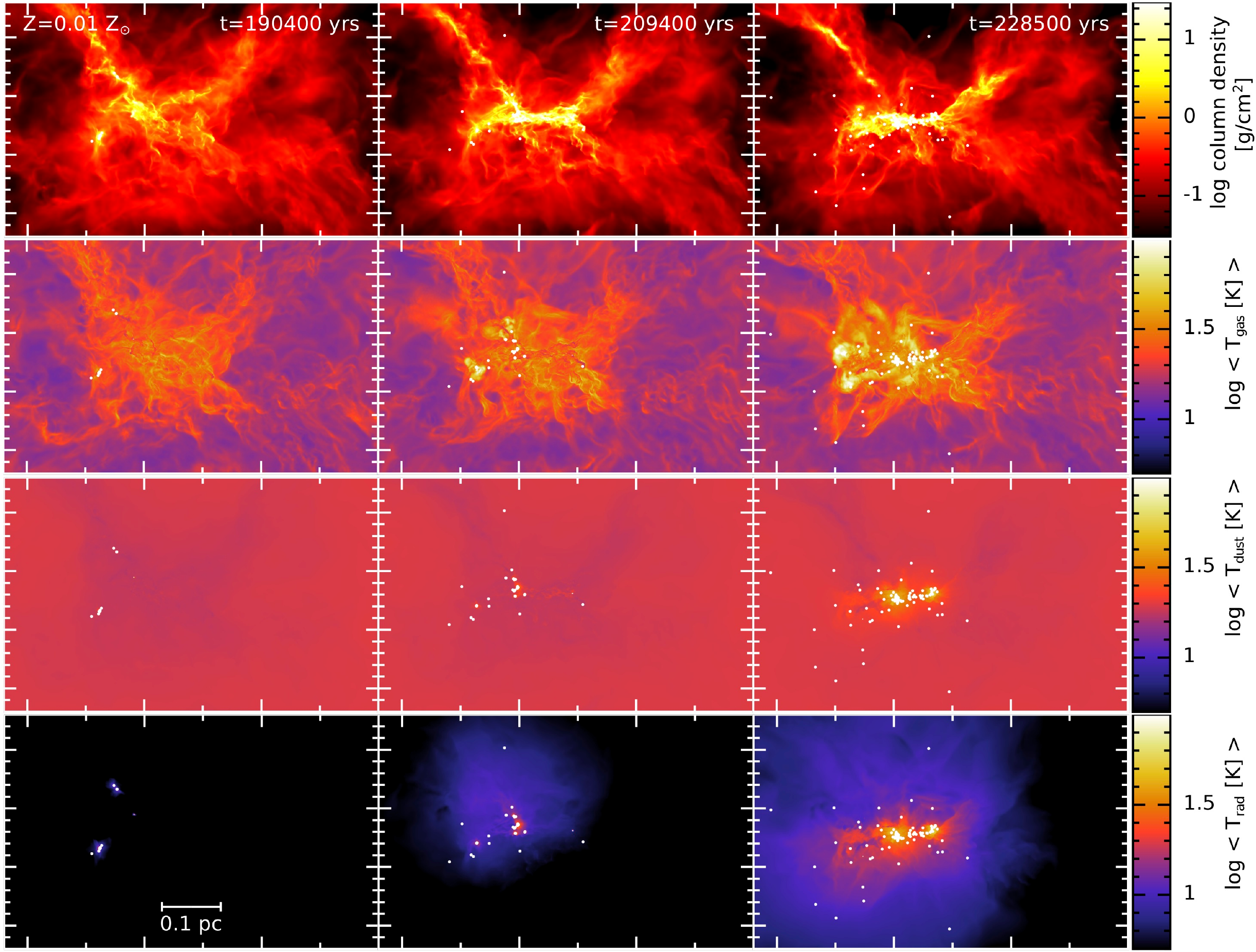} \vspace{0cm}
\caption{Column density and temperature snapshots at three different times ($t=1.00, 1.10, 1.20$~$t_{\rm ff}$ ) for the $z=5$ calculation with a metallicity of 1/100 times solar. From top to bottom, the rows give column density, and the mass-weighted gas, dust, and protostellar radiation temperatures.  The colour scales are logarithmic.  The column density scale covers $0.03-30$~g~cm$^{-3}$, and the temperature scales cover $5-100$K.  The stars and brown dwarfs are plotted using white circles. Gas temperatures in the dense parts of the cloud are much hotter than the dust temperature due to the low metallicity and poor gas-dust coupling.  This figure and caption are comparable to Fig.~2, \citet{Bate2019}.  Animations of the evolution shown in this figure are provided in the online Additional Supporting Information that accompanies this paper.  }
\label{fig:DTZ001}
\end{figure*}

\begin{figure*}
\centering
    \includegraphics[width=17.5cm]{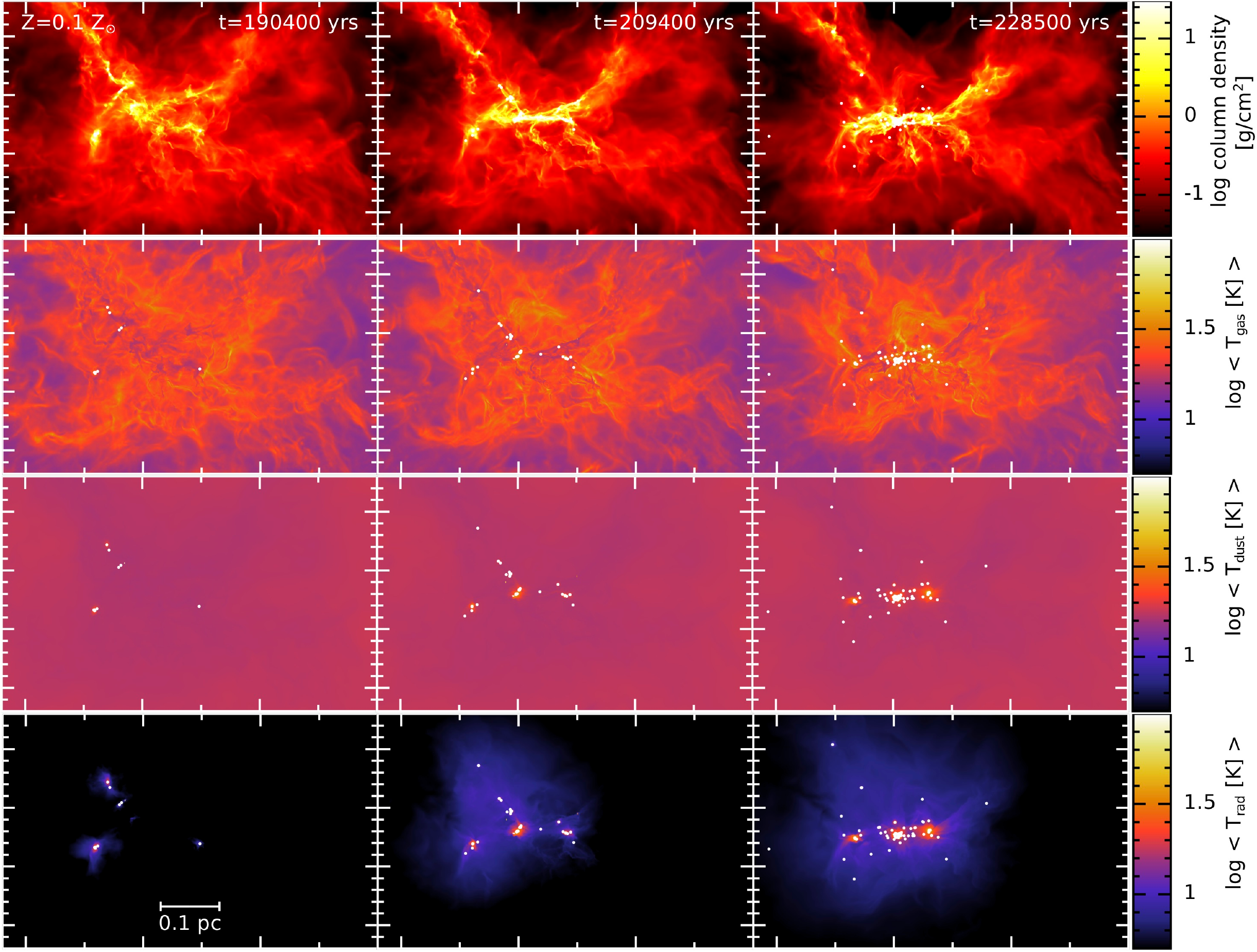} \vspace{0cm}
\caption{Column density and temperature snapshots at three different times ($t=1.00, 1.10, 1.20$~$t_{\rm ff}$ ) for the $z=5$ calculation with a metallicity of 1/10 times solar. From top to bottom, the rows give column density, and the mass-weighted gas, dust, and protostellar radiation temperatures.  The colour scales are logarithmic.  The column density scale covers $0.03-30$~g~cm$^{-3}$, and the temperature scales cover $5-100$K.  The stars and brown dwarfs are plotted using white circles. The gas temperatures in the dense parts of the cloud tend to be hotter than the dust temperature, but less than in the 1/100 metallicity case.  This figure and caption are comparable to Fig.~3, \citet{Bate2019}.  Animations of the evolution shown in this figure are provided in the online Additional Supporting Information that accompanies this paper.  }
\label{fig:DTZ01}
\end{figure*}

\begin{figure*}
\centering
    \includegraphics[width=17.5cm]{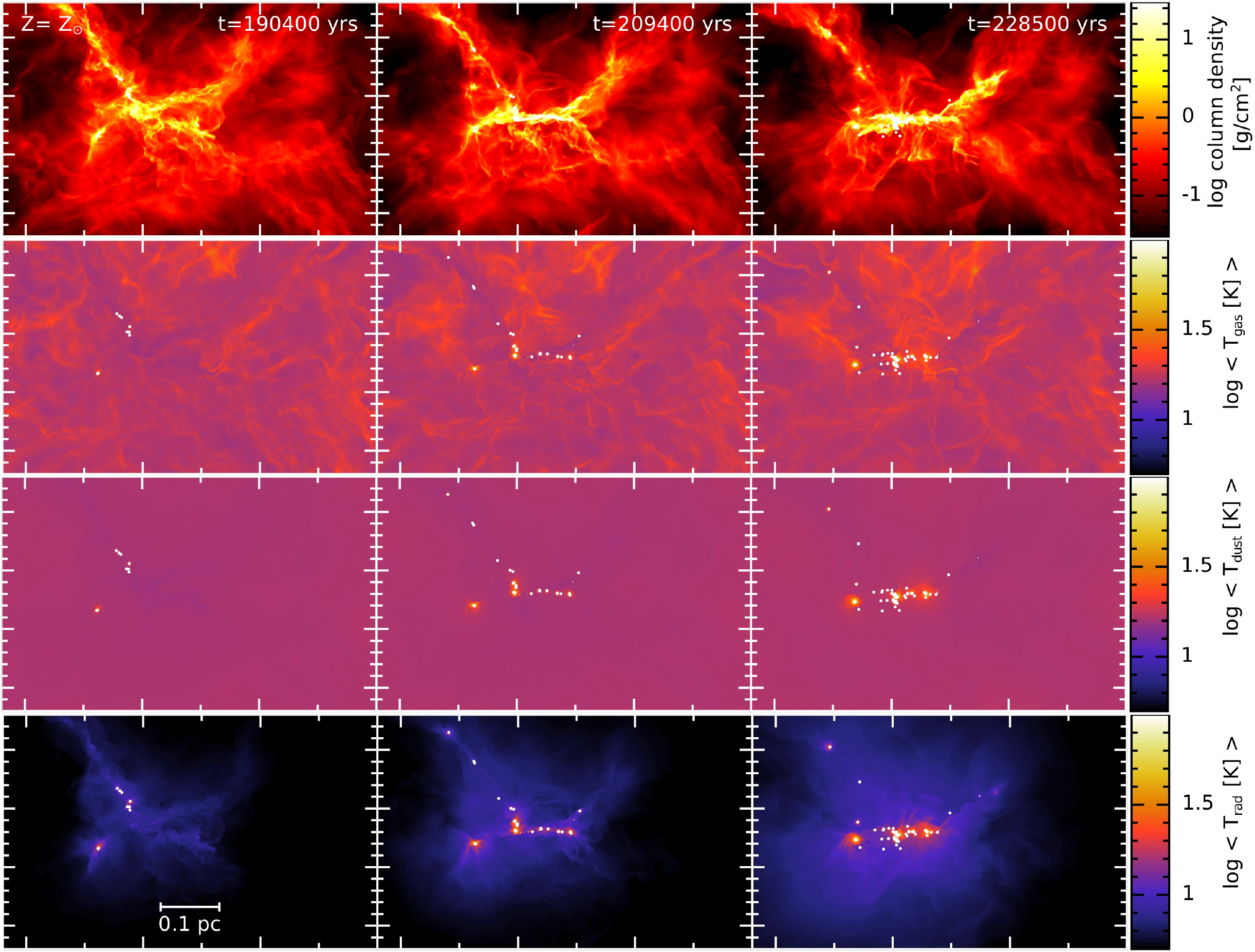} \vspace{0cm}
\caption{Column density and temperature snapshots at three different times ($t=1.00, 1.10, 1.20$~$t_{\rm ff}$ ) for the $z=5$ calculation with solar metallicity. From top to bottom, the rows give column density, and the mass-weighted gas, dust, and protostellar radiation temperatures.  The colour scales are logarithmic.  The column density scale covers $0.03-30$~g~cm$^{-3}$, and the temperature scales cover $5-100$K.  The stars and brown dwarfs are plotted using white circles. The gas temperatures are generally lower than in the low metallicity calculations.  This figure and caption are comparable to Fig.~4, \citet{Bate2019}.  Animations of the evolution shown in this figure are provided in the online Additional Supporting Information that accompanies this paper.  }
\label{fig:DTZ1}
\end{figure*}

\section{Results}
\label{sec:results}

\subsection{Cloud evolution}
\label{sec:clouds}

Each of the three radiation hydrodynamical calculations was evolved to $t=1.20~t_{\rm ff}$ (228,300~yrs).  During this period they produced different numbers of stars and brown dwarfs (modelled using sink particles with accretion radii of 0.5~au).  The initial conditions and the statistical properties of the stars and brown dwarfs produced by each $z=5$ calculation are summarised in Table \ref{table1}, along with the properties of the $z=0$ calculations that were previously published by \citet{Bate2019}.   

At both redshifts there is a clear trend that lower metallicity clouds have converted less gas into stars at the same time.  For the present-day calculations, \citet{Bate2019} showed that this was due to a delay in the star formation that was caused by more metal-poor gas at intermediate densities being warmer (i.e.\ greater thermal pressure support).  Fig.~\ref{massnumber} shows how the star formation rates evolve with time for the $z=5$ calculations, both in terms of the number of stars and brown dwarfs and their total mass.  As was the case in the $z=0$ calculations, it is clear from the left panel of these graphs that the trend of lower total stellar mass with lower metallicity at the end of the calculations ($t=1.20~t_{\rm ff}$) arises primarily because the star formation is delayed with lower metallicity.  After $t=1.08~t_{\rm ff}$ (205,500~yrs), the star formation rates (the slopes of the lines in the left panel) are almost identical.  But there is a delay in the star formation getting started that increases for lower metallicities.  At the end of the calculations, almost 1/7 of the initial mass has been transformed into stars in the solar metallicity calculation, but in the lowest metallicity calculation only 1/10 of the mass has been converted to stars.  Note that for the same metallicity, the $z=5$ calculations have generally converted less gas into stars after the same amount of time than the $z=0$ calculations -- the amount of mass is similar for $Z=0.01~{\rm Z}_\odot$ (actually 5 per cent higher), but substantially less for $Z\geq 0.1~{\rm Z}_\odot$.  As we will see below, this is because the gas temperatures at the same metallicity tend to be higher for $z=5$ than for $z=0$ due to the enhanced CMBR.

However, while the trend of the total stellar mass with metallicity is the same for the $z=0$ and $z=5$ calculations, the dependence of the number of objects with metallicity actually reverses for $z=5$ compared to $z=0$.  Whereas the present-day star formation calculations of \citet{Bate2019} produced more objects at higher metallicities at $t=1.20~t_{\rm ff}$, in the $z=5$ calculations higher metallicity results in fewer objects being formed.  The greater numbers of objects at higher metallicity in the present-day star formation calculations is due to lower gas temperatures caused by the combination of greater extinction shielding the inner parts of the cloud from the ISRF and enhanced dust cooling.  However, the latter effect starts to saturate at very high metallicity ($Z \approx 1-3~Z_\odot$) when the cloud is highly optically thick and begins to trap its own infrared radiation, so that at $z=0$ the calculations with $Z=Z_\odot$ and $Z=3~Z_\odot$ produced similar numbers of objects: 255 \& 258, respectively (Table \ref{table1}; \citealt{Bate2019}).  By contrast, at redshift $z=5$ the ability of the gas to cool is diminished by the higher CMBR temperature.  Moreover, at higher temperatures the clouds are more optically thick to their own radiation (i.e.\ the Rossland mean opacity increases with temperature).  These effects combine to reverse the present-day metallicity trend.  The $Z=0.01~{\rm Z}_\odot$ and $Z=0.1~{\rm Z}_\odot$ calculations produce reasonably similar numbers of objects (132 and 114, respectively), but the $Z={\rm Z}_\odot$ produces only 67 objects.

\begin{figure*}
\centering
    \includegraphics[height=9cm]{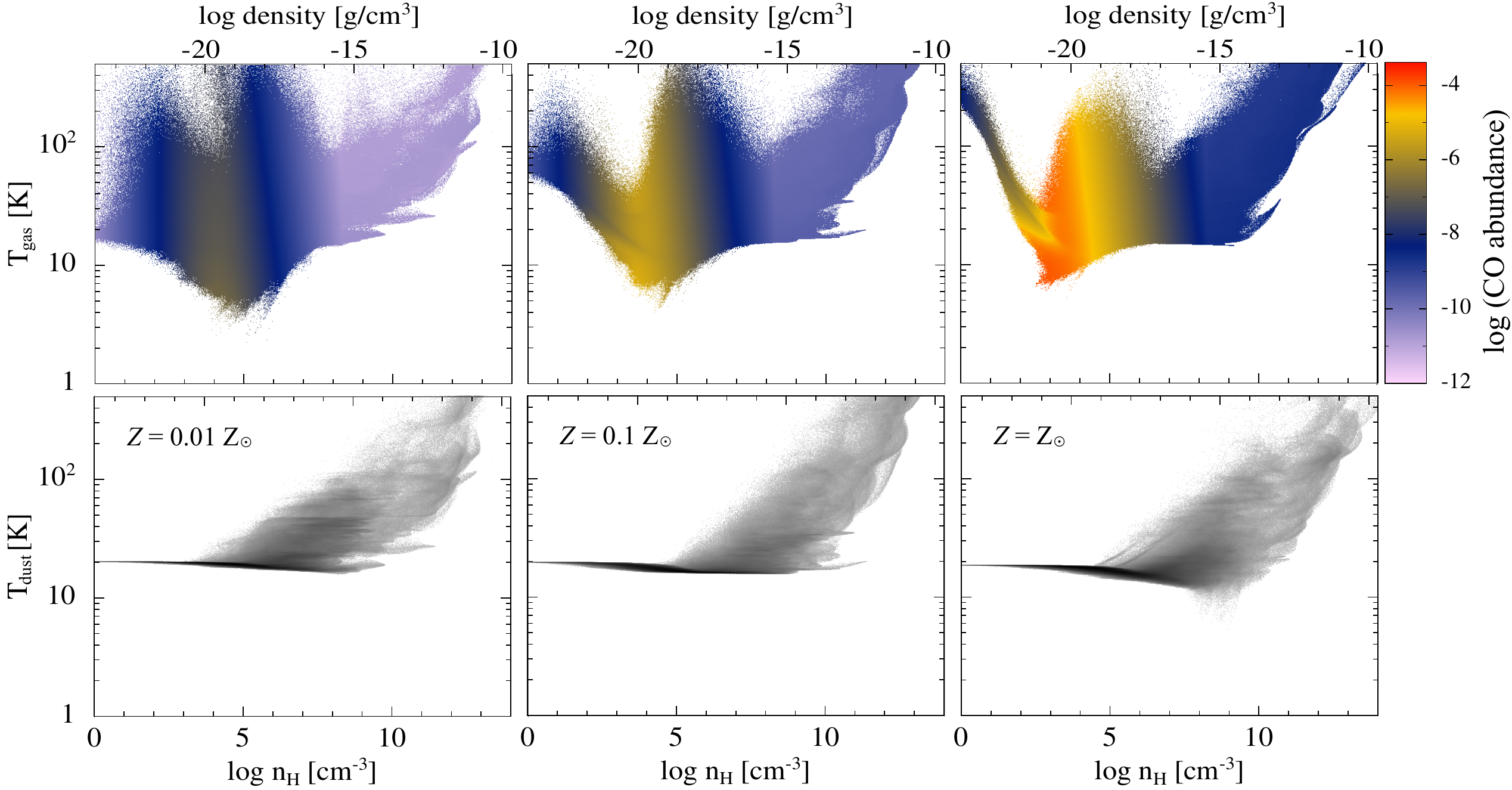} \vspace{0cm} 
\caption{Phase diagrams of temperature vs gas density at the end of each of the $z=5$ calculations ($t=1.20$~$t_{\rm ff}$). The upper row of panels gives the gas temperature, while the lower row gives the dust temperature. The colour scale in the upper panels gives the mean abundance of gas phase CO relative to hydrogen at that temperature and density. Note that the calculations include a prescription for the freeze out of CO onto dust grains, but they do not treat thermal desorption of CO from dust grains at $T_{\rm dust}\gsim 20$~K.  In the dust temperature plots, the grey scale is proportional to the logarithm of the number of fluid elements (i.e., darker regions contain more SPH particles).  The low-density gas ($n_{\rm H}\lsim 10^{8}~{\rm cm}^{-3}$) tends to be warmer at low metallicity than with high metallicity due to the poor cooling, but the high-density gas ($n_{\rm H}\gsim 10^{10}~{\rm cm}^{-3}$) tends to be cooler at low metallicity due to the reduced optical depths. The dust is almost all warmer than 10~K, as opposed to the present-day star formation calculations of \citet{Bate2019} in which dust temperatures as low as $5-6$~K were reached with metallicity $Z \gsim {\rm Z}_\odot$. Since dust cooling is crucial for gas cooling at high densities and metallicities, this leads to significantly greater gas temperatures for $Z \gsim 0.1 ~{\rm Z}_\odot$ at $z=5$ than at $z=0$.  This figure and caption are comparable to Fig.~7, \citet{Bate2019}.  }
\label{fig:phase}
\end{figure*}

The mean and median masses of the objects that are produced in each of the clouds are provided in columns 7--9 of Table \ref{table1}. \cite{Bate2019} found that these statistical properties of the stars/brown dwarfs produced by the calculations at $z=0$ were statistically indistinguishable over the full range of metallicities from $Z=0.01-3~{\rm Z}_\odot$.  Although at a given time more mass was converted into stars at higher metallicity, more objects were produced from that gas leading to a protostellar mass function that was independent of the metallicity.  However, at $z=5$ although at a given time more mass is again converted into stars at higher metallicity (left panel of Fig.~\ref{massnumber}), the number of objects produce by that gas consistently decreases with increasing metallicity (centre and right panels of Fig.~\ref{massnumber}).  The result is that the typical stellar mass increases with increasing metallicity  (Table \ref{table1}).   There is only a small difference between the stellar populations produced by the $Z=0.01~{\rm Z}_\odot$ and $Z=0.1~{\rm Z}_\odot$ calculations, but the characteristic mass doubles (either the mean or median mass) between the $Z=0.1~{\rm Z}_\odot$ and $Z={\rm Z}_\odot$ calculations.

In all of the calculations there are some protostellar mergers (see the last column of Table \ref{table1}).  These occur when two sink particles pass within $6~{\rm R}_\odot$ of each other.  This merger radius was chosen because low-mass protostars that accrete at high rates ($\dot{M}_* \sim 10^{-6} - 10^{-5}$~M$_\odot$~yr$^{-1}$) are believed to have radii of $2-3~{\rm R}_\odot$ \citep[e.g.][]{Larson1969,HosOmu2009}.  At both redshifts the average number of stellar mergers per star consistently increases with decreasing metallicity.  This is due to the lower opacities at lower metallicity that allow faster cooling of very high-density gas and, therefore, more small-scale fragmentation \citep[c.f.][]{Bate2019}.

In the right panel of Fig.~\ref{massnumber}, we plot the number of stars and brown dwarfs versus their total mass.  There appears to be a change in the nature of the star formation in the solar-metallicity calculation when $\approx 45$~M$_\odot$ of gas has been converted into stars in that beyond this point fewer new objects are created per unit mass of gas.  This is reminiscent of the calculations of \citet*{KruKleMcK2011} in which the star-forming clouds were so dense and the protostars so close together that the radiative feedback from the young protostars heating the high-density gas inhibited the formation of new protostars, resulting in a top-heavy mass function.  The same effect seems to be happening in the solar-metallicity calculation here, but for different reasons: the combination of the inherently hotter clouds due to the enhanced CMBR and more metal-rich high-density gas trapping radiation and being less able to cool is inhibiting the formation of new protostars.  Thus, the collapsing gas primarily gets accreted by existing protostars, boosting their mass, rather than producing new objects.

The reason for the delay of the star formation at low metallicities is that the intermediate-density, metal-poor gas is hotter.  In Figs.~\ref{fig:DTZ001} to \ref{fig:DTZ1} we provide images of the column densities and mass-weighed temperatures from the three $z=5$ calculations.  The times ($t=1.00, 1.10, 1.20$~$t_{\rm ff}$) have been chosen to cover the main periods of star formation.  In each figure, the second, third, and fourth rows give the separate gas, dust, and radiation temperature distributions, respectively.  The radiation temperature is that of the radiation field from the collapsing gas and dust, modelled using flux-limited diffusion; the ISRF that also heats the dust is treated separately.

For $z=0$, \citet{Bate2019} found a clear progression that both gas and dust temperatures were hotter at lower metallicities for $Z \le {\rm Z}_\odot$.  For example, in the $z=0$ calculations at higher metallicities the inner parts of the clouds were more shielding from the ISRF, resulting in typical dust temperatures on large spatial scales that decreased with increasing metallicity.  This is not the case at $z=5$.  The dust temperatures are much more uniform due to the heating from the hotter CMBR (which is long wavelength radiation that is much more able to penetrate the clouds than the high frequency components of the ISRF). This is further illustrated in lower panels of Fig.~\ref{fig:phase} which shows phase diagrams of the gas and dust temperatures functions of density, that can be directly compared to those in \citet{Bate2019}.  The upper panels of Fig.~\ref{fig:phase} provide the gas-phase CO abundance relative to hydrogen using a colour scale.  Note that the chemical model treats the freeze out of CO on to dust grains and desorption of CO by cosmic rays, but it does not treat the thermal desorption that occurs at dust temperatures above 20~K (e.g. in protostellar discs).  This is neglected because the only role of the chemistry in the calculations is to provide realistic gas temperatures and the primary coolant at the densities above which CO freeze out becomes important is usually the dust rather than the CO.
In the lower panels of Fig.~\ref{fig:phase} the grey scale is proportional to the logarithm of the amount of dust at each temperature and density.
In the $z=5$ calculations, the dust temperatures at low and intermediate densities are $T_{\rm dust} \approx 15-20$~K and even in the solar metallicity calculation the coldest dust barely reaches down to $T_{\rm dust} \approx 10$~K.  By contrast, at $z=0$ the dust was able to cool to much lower temperatures of $T_{\rm dust} \approx 10$~K at $Z = 0.01~{\rm Z}_\odot$, $T_{\rm dust} \approx 8$~K at $Z = 0.1~{\rm Z}_\odot$, $T_{\rm dust} \approx 6$~K at $Z \geq {\rm Z}_\odot$ \citep{Bate2019}. 

The inability of the dust to cool in the $z=5$ calculations means that the gas is also unable to cool as it did in the $z=0$ calculations.  At low gas densities ($n_{\rm H} \lsim 10^3$~cm$^{-3}$) the spread of gas temperatures (Fig.~\ref{fig:phase}) is reasonably similar in the $z=0$ and $z=5$ calculations \citep[c.f.\ Fig.~7 of][]{Bate2019}.  At both redshifts, at intermediate gas densities ($n_{\rm H} \approx 10^4-10^7$~cm$^{-3}$) the hotter gas tends to be a bit hotter at lower metallicity (see Fig.~\ref{fig:phase}) because metal-poor gas is less able to cool than metal-rich gas due to the reduction of the atomic and molecular abundances (that dominate the cooling at low densities).  When shocks and other compressive motions in the clouds increase the thermal energy, the low-metallicity gas cannot radiate this heat away as easily and, thus, the gas is significantly hotter.  This hotter gas gives greater gas pressure support against gravity to the metal-poor clouds and delays the collapse of the low metallicity clouds, as seen in Fig.~\ref{massnumber}.  Comparing the temperature ranges of gas at intermediate densities at the same metallicity but different redshift, the hottest gas is very similar, but the coldest gas is colder in the $z=0$ calculations than in the $z=5$ calculations.

At higher densities this difference in the lower range of gas temperatures between calculations with the same metallicity but at different redshifts becomes even more significant, particularly at high metallicity.
The cooling of high-density molecular gas with $Z \gsim 0.01~Z_\odot$ is primarily dominated by dust continuum cooling; gas and dust are thermally coupled by collisions.  Dust and gas temperatures become similar above number densities
\begin{equation}
n_{\rm H} \approx 10^6 \left( \frac{Z}{Z_\odot} \right)^{-1}~\mbox{\rm cm}^{-3},
\end{equation}
as can be seen in Fig~\ref{fig:phase} \citep[see also][]{Omukai2000,GloCla2012c,Bate2019}.  In the calculations of present-day star formation ($z=0$), the cooler dust temperatures at higher metallicities produced substantially lower gas temperatures at higher metallicity.  This trend with increasing metallicity is almost completely absent in the $z=5$ calculations.  In Fig.~\ref{fig:phase} at high densities ($n_{\rm H} \geq 10^7$~cm$^{-3}$) there is very little difference in the spread in $T_{\rm gas}$-n$_{\rm H}$ for different metallicities because the dust temperatures are very similar and the much warmer dust is not as effective at cooling the gas.

It is this comparatively ineffective dust cooling that is fundamentally responsible for the differences in the stellar properties between the $z=0$ and $z=5$ calculations.  For the same metallicity ($Z \geq 0.1~{\rm Z}_\odot$), the dust embedded in the high-density gas is substantially warmer in the $z=5$ calculations than in the $z=0$ calculations.  Furthermore, the magnitude of this difference increases with increasing metallicity because at $z=0$ the high-density gas becomes colder and colder at higher metallicity \citep[see Fig.~7 of][]{Bate2019}.  Thus, the gas is less prone to fragment into protostars in the $z=5$ calculations, producing fewer protostars for the same amount of gas that is converted into stars and, therefore, the characteristic stellar masses are higher in the $z=5$ calculations than in the $z=0$ calculations with the same metallicity.  Since the $z=0$ calculations produced protostars with a characteristic mass that was independent of metallicity, this change in the thermodynamics means that the $z=5$ calculations instead produce a characteristic stellar mass that increases with increasing metallicity.

As the star formation proceeds, the radiation fields generated by the gas collapsing into the gravitational potential wells of the protostars become stronger (bottom rows of Figs.~\ref{fig:DTZ001} to \ref{fig:DTZ1}).  These primarily heat the gas and dust with the highest densities.  The effect of this protostellar feedback is much less obvious on the gas and dust temperatures in Figs.~\ref{fig:DTZ001} to \ref{fig:DTZ1} than it was in the equivalent figures of \citet{Bate2019} at $z=0$ because of the generally higher dust and gas temperatures due to the hotter CMBR (i.e.\ the protostellar radiation fields tend to be swamped on scales $\gsim 0.02$~pc by the greater overall level of radiation from the CMBR).

\begin{figure}
\centering 
    \includegraphics[height=5.0cm]{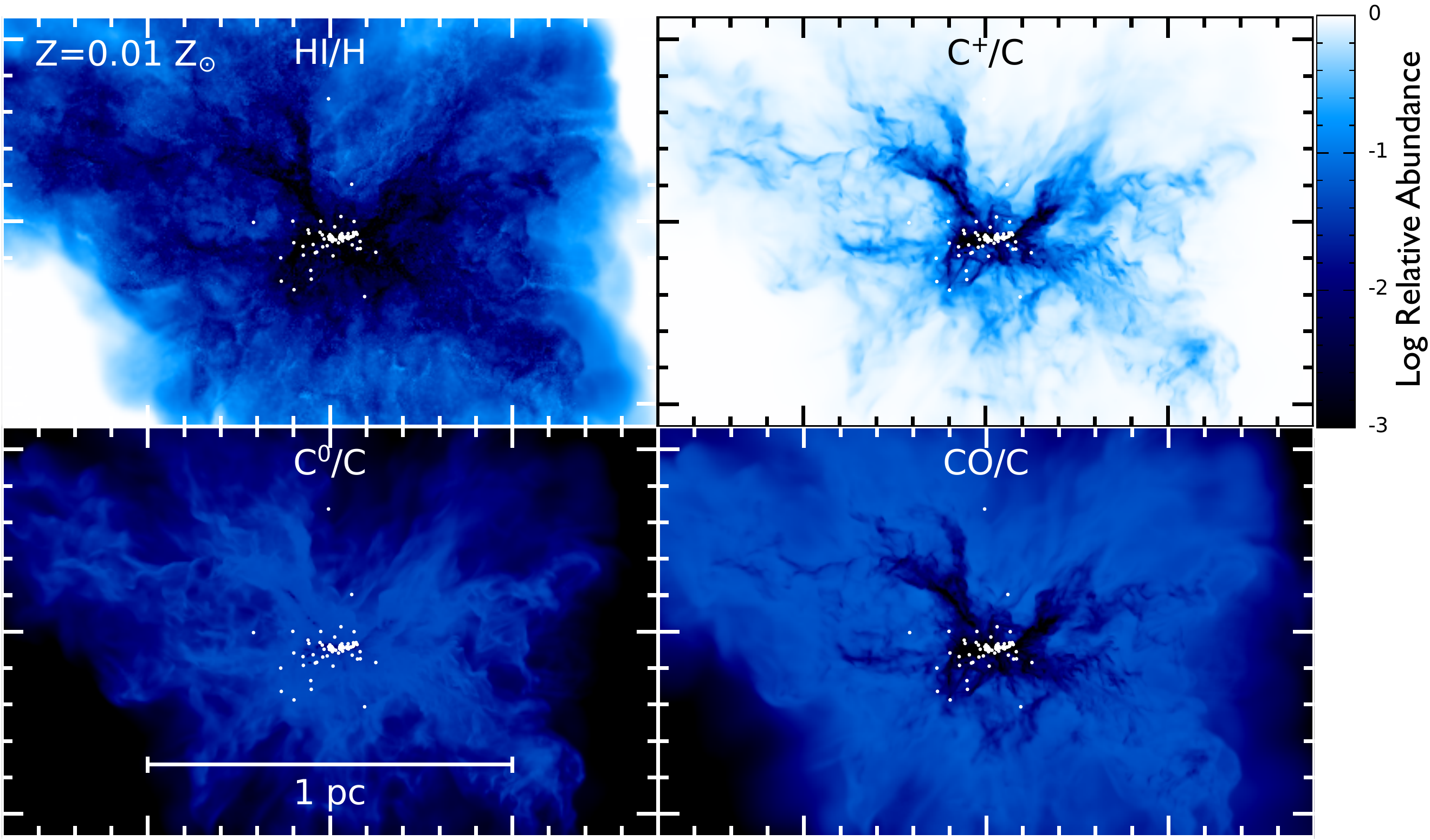} \vspace{0cm} \vspace{0cm}
    \includegraphics[height=5.0cm]{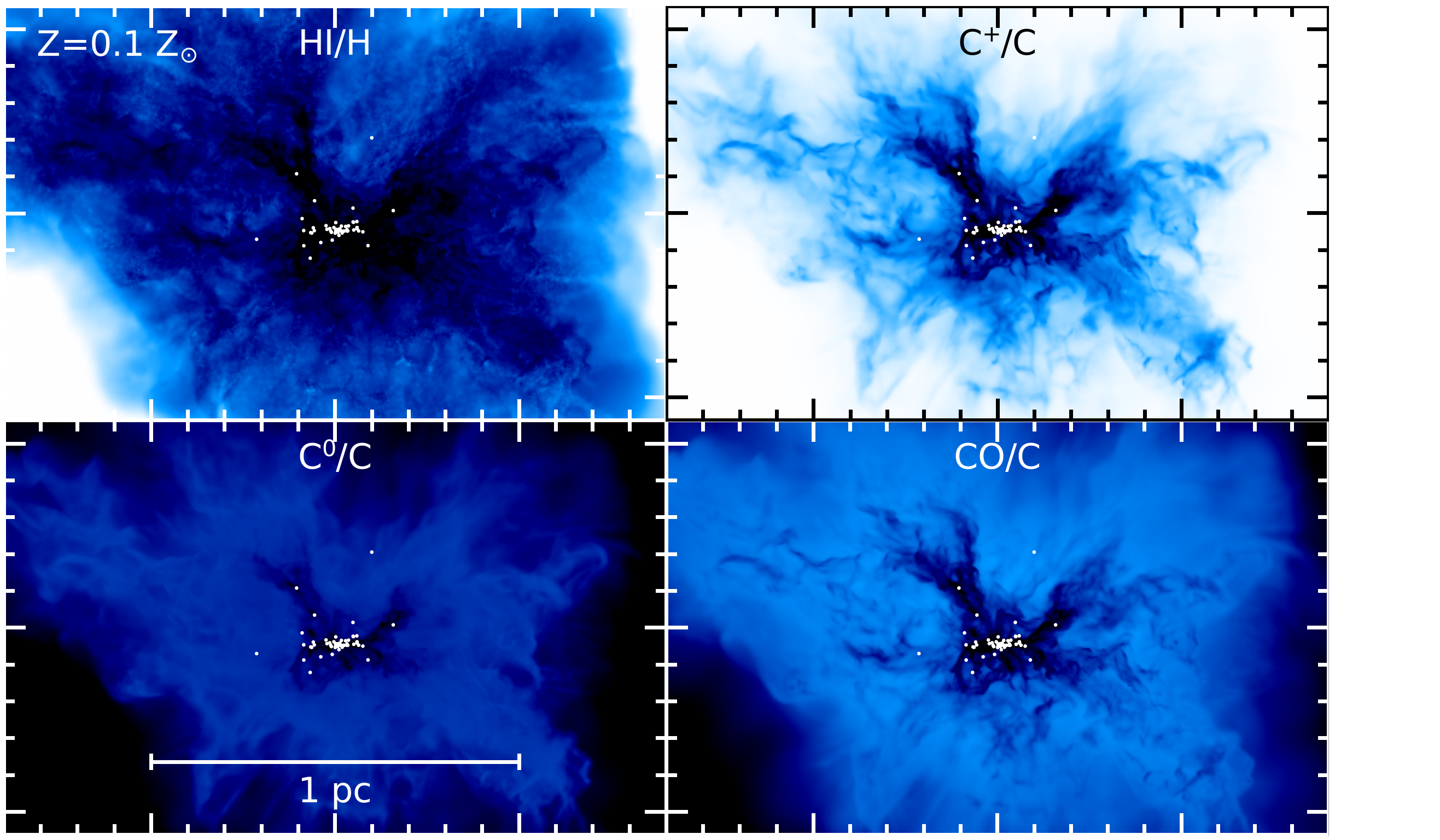} \vspace{0cm} \vspace{0.cm}
    \includegraphics[height=5.0cm]{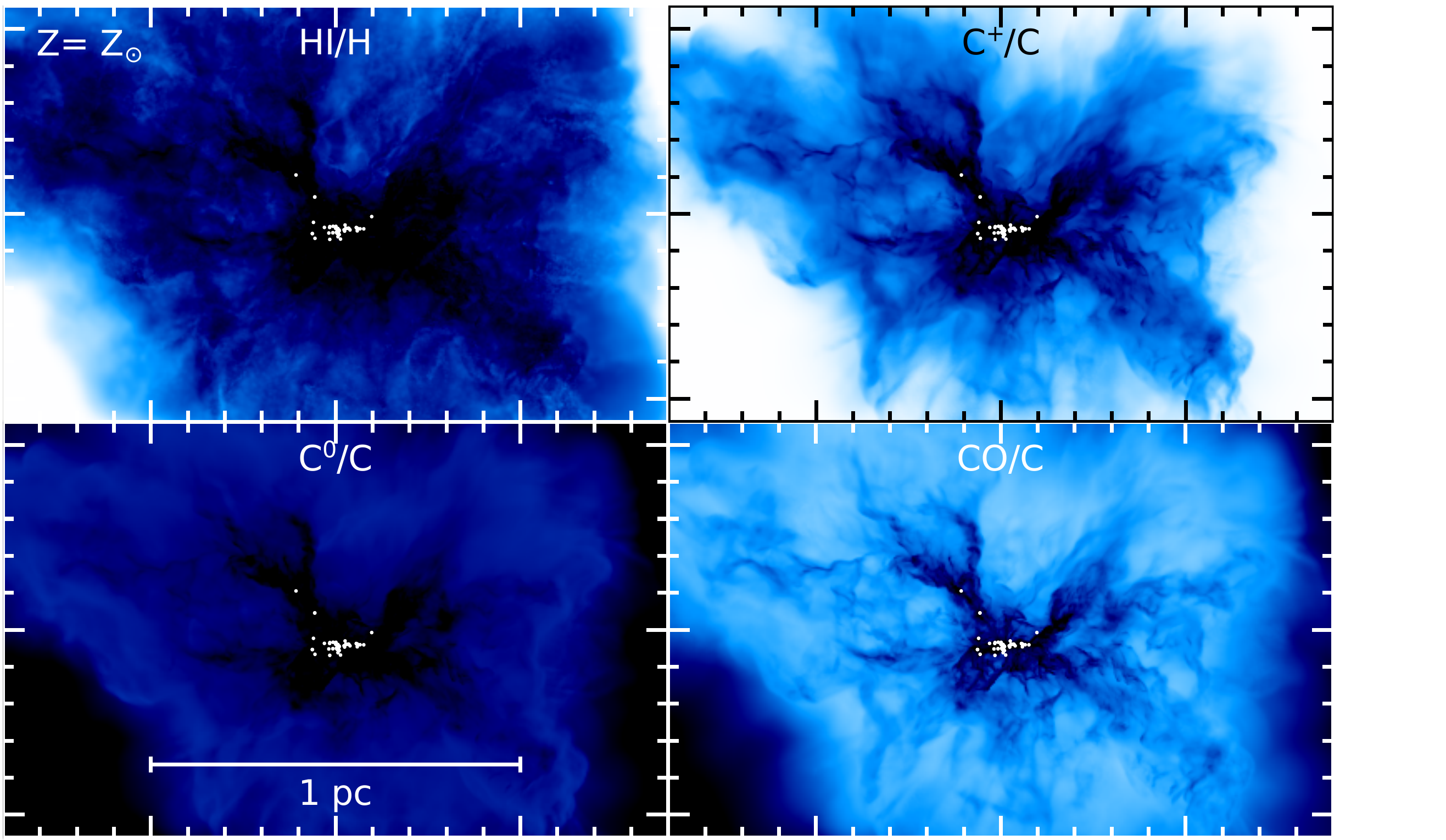} \vspace{0cm}
\caption{Snapshots of the mass-weighted abundances of atomic hydrogen, ionised carbon, neutral carbon, and CO at the end of each of the three calculations with differing metallicity ($t=1.20$~$t_{\rm ff}$). There are three groups of panels, one for each metallicity of 1/100, 1/10, and 1 times solar. For each group, the top left panel gives the relative abundance HI/H, the top-right panel the relative abundance C$^+$/C, the lower-left panel the relative abundance C$^0$/C, and the lower-right panel the relative abundance of gas phase CO to the total carbon abundance.  The stars and brown dwarfs are plotted using white dots. The colour scale of relative abundances is logarithmic ranging from $10^{-3}$ to unity.  Generally, the outer parts of the clouds contain atomic hydrogen and C$^+$.  The inner parts of the cloud are primarily composed of H$_2$ and carbon is in the form of CO.  At the highest densities, CO freezes out onto grains (hence the drop in gas-phase abundance).  At lower metallicities, there is a greater fraction of atomic hydrogen and more of the carbon is in C$^+$ because there is less extinction of the ISRF by dust and more self-shielding of H$_2$. This figure and caption are comparable to Fig.~6, \citet{Bate2019}.  }
\label{fig:chemistrysnaps}
\end{figure}

\begin{figure}
\centering \vspace{-0.5cm}
\includegraphics[width=9cm]{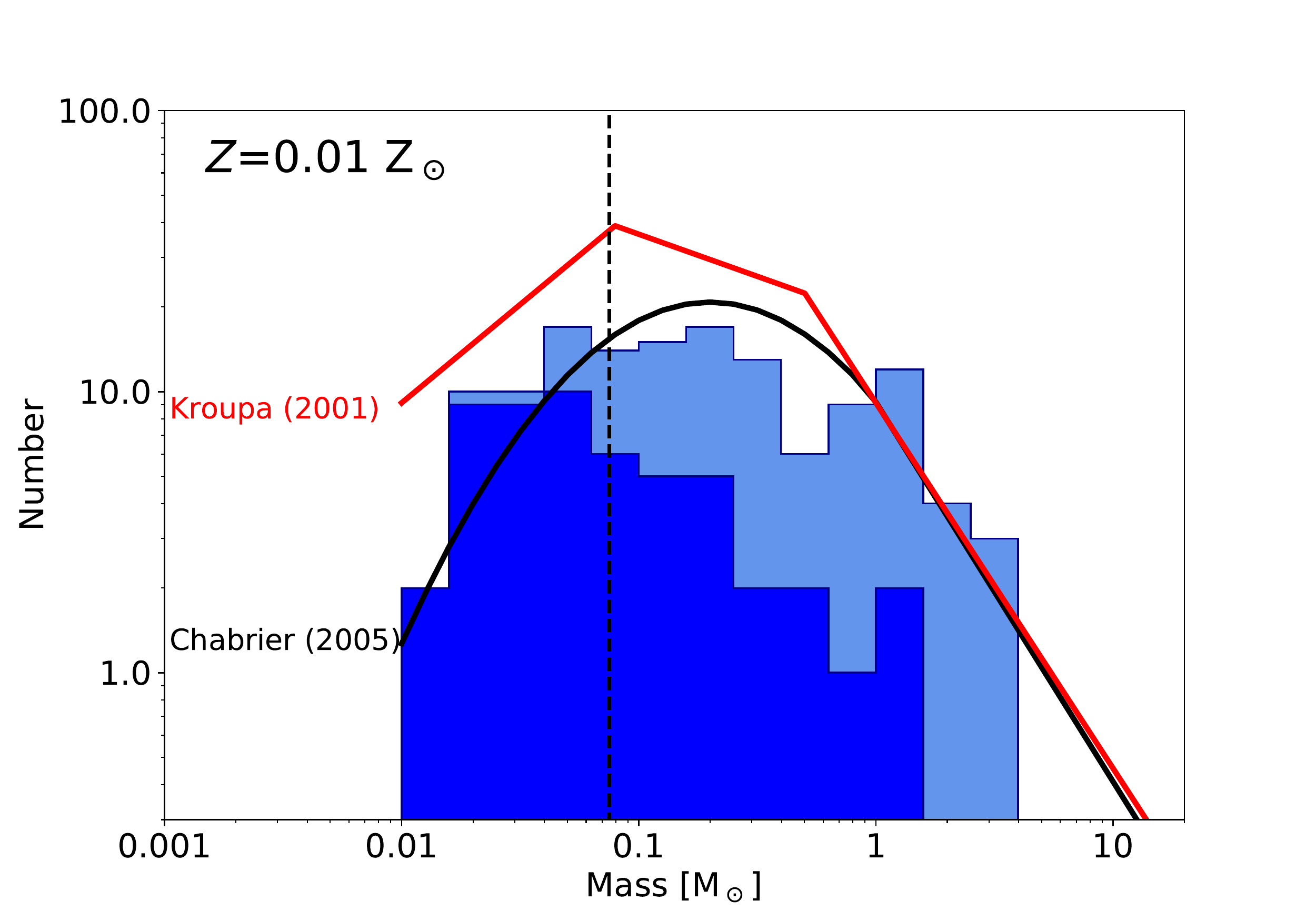}  
\includegraphics[width=9cm]{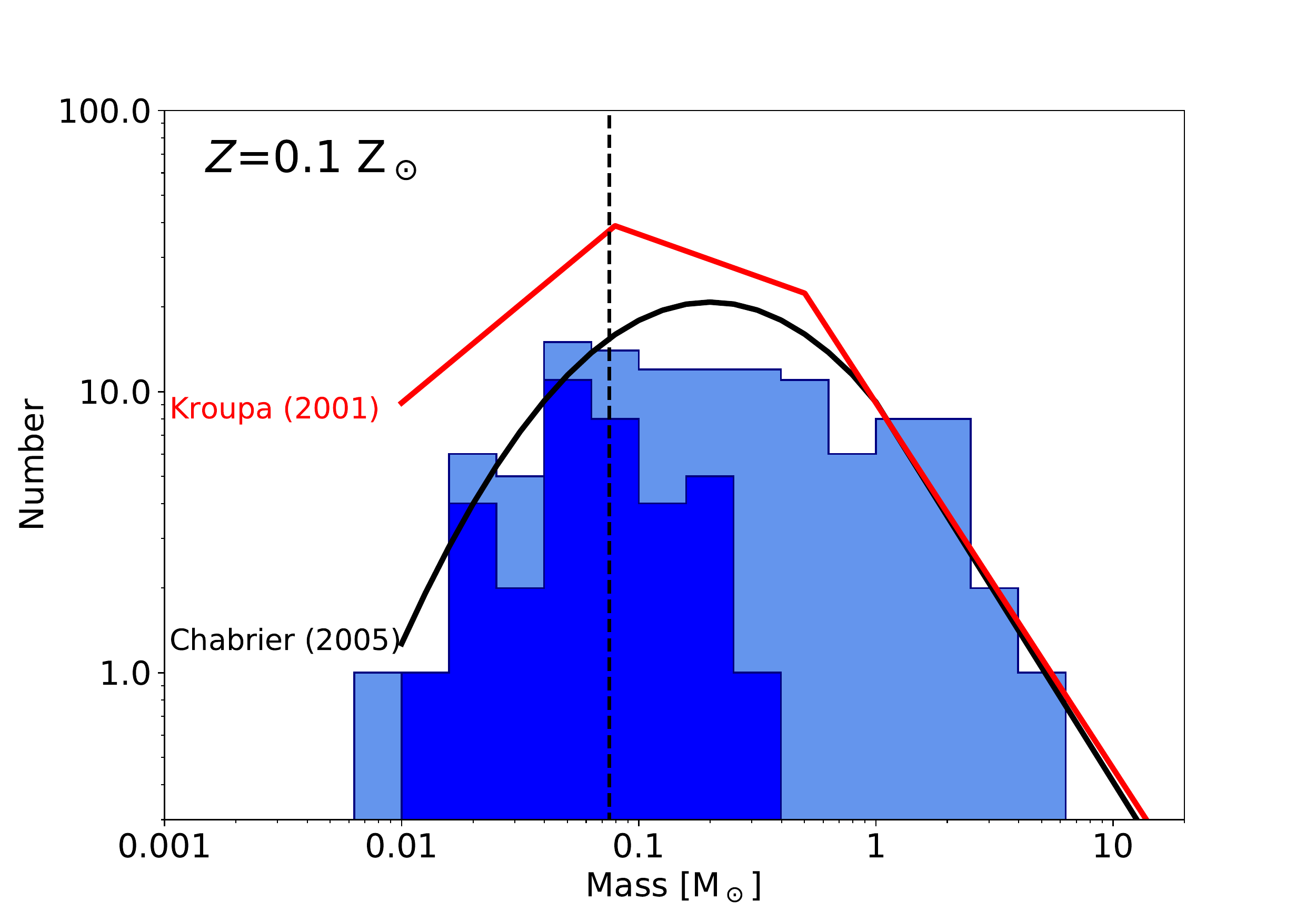}  
\includegraphics[width=9cm]{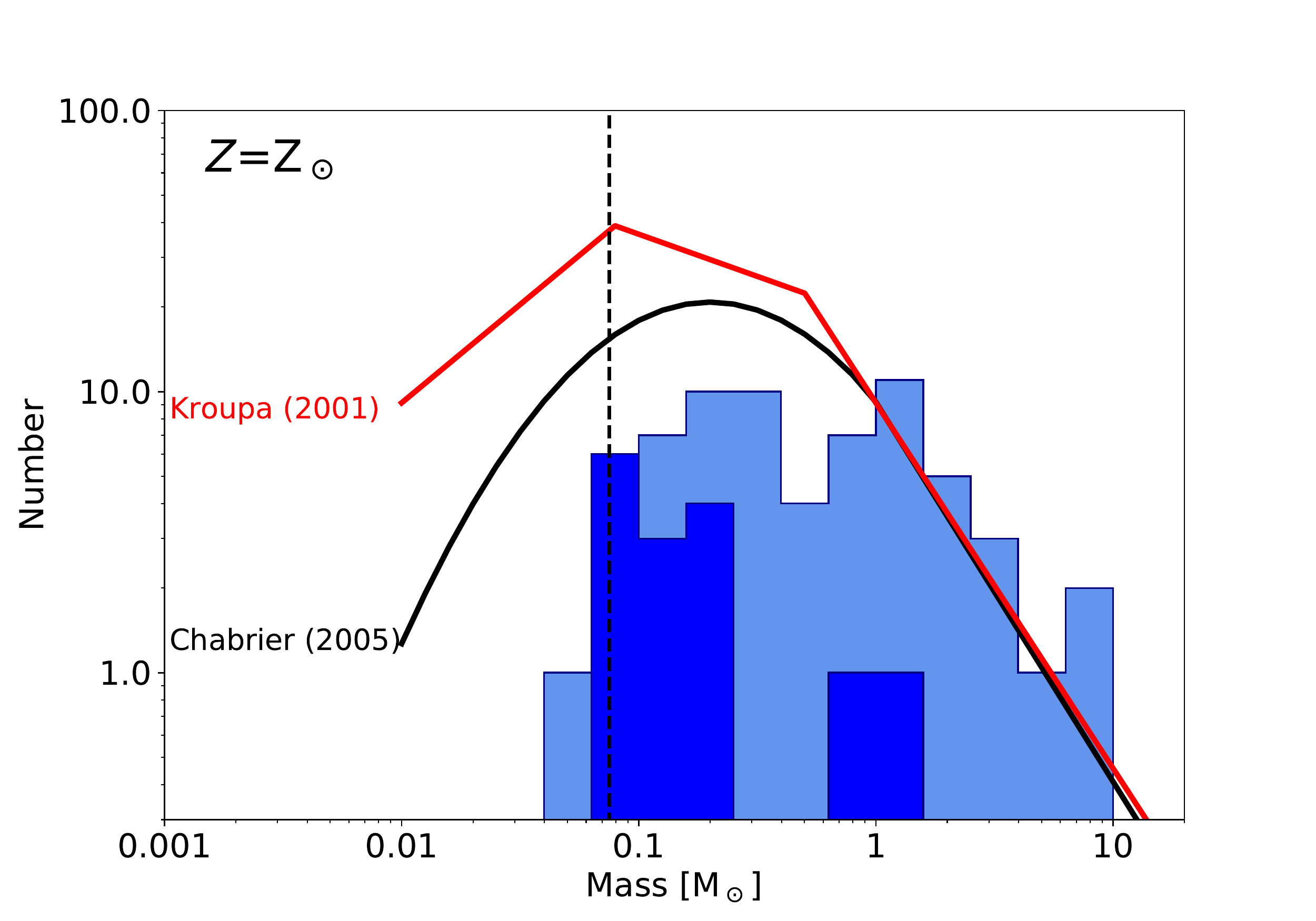}
\caption{Histograms giving the distributions of the masses of the stars and brown dwarfs produced by the three radiation hydrodynamical calculations, each at $t=1.20t_{\rm ff}$.  The dark blue histograms are used to denote those objects that have stopped accreting (defined as accreting at a rate of less than $10^{-7}$~M$_\odot$~yr$^{-1}$), while those objects that were still accreting when the calculations were stopped are plotted using light blue.  The \citet{Kroupa2001} and \citet{Chabrier2005} parameterisations of the IMF are also plotted. The vertical dashed line marks the stellar/brown dwarf boundary.  The two low-metallicity calculations produce mass functions that are in reasonable agreement with the \citet{Chabrier2005} fit to the observed IMF for individual objects, but the solar metallicity calculation is noticeably deficient in brown dwarfs and low-mass stars.  This figure and caption are comparable to Fig.~8, \citet{Bate2019}.   }
\label{fig:IMF}
\end{figure}

\subsection{Chemistry}

The present-day star formation calculations of \cite{Bate2019} were the first to combine a model of the diffuse ISM with radiative transfer to treat the radiation produced by the collapsing gas.  The calculations keep track of whether the hydrogen is in atomic or molecular form, and whether carbon is in the form of ionised carbon, C$^+$, neutral carbon, C$^0$, or molecular in the form of CO.  There is also a model for the freezing out of CO onto dust grains \citep[see][]{BatKet2015}.  Modelling the chemistry is important because the formation of molecular hydrogen is a potential source of heating, while carbon atoms and molecules are important coolants.

Following \citet{Bate2019}, in Fig.~\ref{fig:chemistrysnaps} we provide snapshots of the chemical make up of the clouds at the end of each of the three $z=5$ calculations.  The chemistry varies most in the low-density gas and in the outer parts of the clouds, so we show the large scales.  The carbon abundances are scaled relative to the total carbon abundance in each calculation (i.e. the absolute carbon abundance is 10 times lower in the $Z=0.1~{\rm Z}_\odot$ calculation than in the calculation at solar metallicity).  

If the panels in Fig.~\ref{fig:chemistrysnaps} are compared with the corresponding panels in Figure 6 of \citet{Bate2019}, very little difference is seen for the same metallicity.  This is because it is the higher energy part of the ISRF that dictates the chemistry of the low-density gas, not the CMBR at $z=0-5$.  In both sets of calculations, the hydrogen deep within the clouds is almost entirely molecular.  The atomic hydrogen abundance is only high in the outer parts of the clouds where there is low extinction of the ISRF by dust and less self-shielding provided by the molecular hydrogen. Although the atomic hydrogen abundance within the cloud is a little higher at low metallicity, even with $Z=0.01~{\rm Z}_\odot$ the hydrogen is 97.8 percent molecular at the end of the calculation.

Similarly, carbon is almost entirely in the form of C$^+$ in the outer parts of the clouds, because it is exposed to the ISRF, while deep within the clouds the main gas phase form of carbon is CO.  Neutral atomic carbon is found at intermediate depths, but it is never very abundant.  The C$^+$ fraction has a greater dependence on metallicity than the fraction of atomic hydrogen.

\subsection{The statistical properties of the stellar populations}

In this section, we compare the statistical properties of stars and brown dwarfs produced by the three calculations. We also compare our results with the those from the present-day star formation calculations of \citet{Bate2019}.  As in \citet{Bate2019}, we study the stellar mass distributions, stellar multiplicities, the distributions of separations of multiple stellar systems, and the mass ratio distributions of binary systems.  In earlier papers performing calculations of a similar scale \citep[e.g.][]{Bate2009a,Bate2012,Bate2014}, other statistics such as the orbital eccentricity distributions of multiple systems, triple system orbits, stellar accretion histories and kinematics, and stellar closest encounters were also examined.  We do not present data on these properties here, however, that we find no evidence that these other properties vary significantly with metallicity or redshift between $z=0$ and $z=5$.

\subsubsection{The initial mass function}

In Fig.~\ref{fig:IMF}, we compare the differential mass functions at the end of each of the three radiation hydrodynamical calculations with different metallicities. We compare them with parameterisations of the observed Galactic IMF, given by \cite{Chabrier2005}, and \cite{Kroupa2001}.  There is a clear difference between the mass distributions, with the solar metallicity case producing stars with a higher typical mass than the two sub-solar metallicity calculations and very few brown dwarfs. 

\begin{figure}
\centering
    \includegraphics[width=8.5cm]{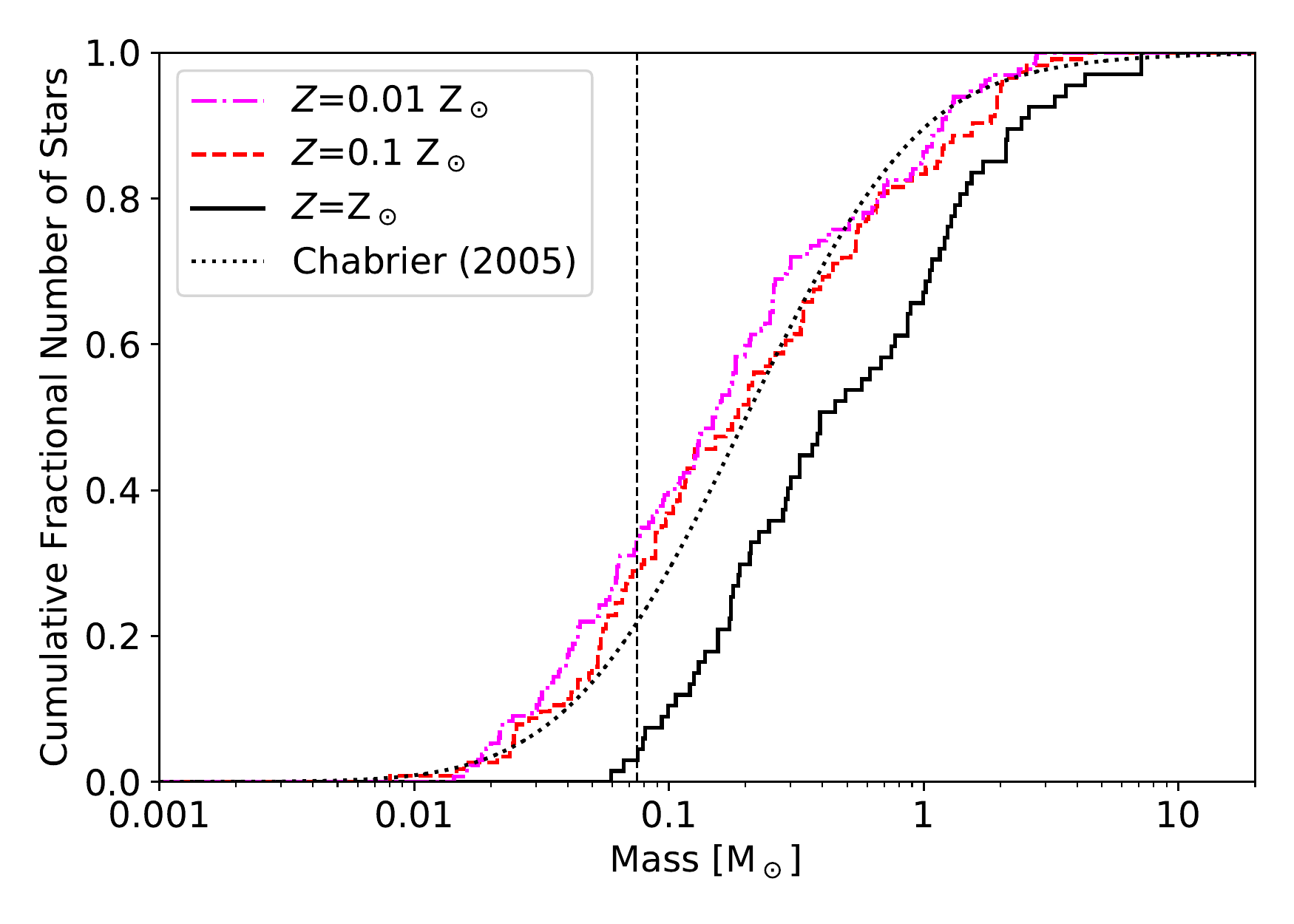} \vspace{-0.5cm}
\caption{The cumulative stellar/brown dwarf mass distributions produced by the three radiation hydrodynamical calculations with metallicities of $Z=0.01~{\rm Z}_\odot$ (magenta dot-dashed line), $Z=0.1~{\rm Z}_\odot$ (red dashed line), and $Z={\rm Z}_\odot$ (black solid line).  We also plot the Chabrier (2005) IMF (black dotted line).  The vertical dashed line marks the stellar/brown dwarf boundary.  The form of the stellar mass distribution is similar for all three calculations, but the median mass increases with metallicity, particularly for the solar metallicity calculation which produces no low-mass ($M_* \lsim 0.05$~M$_\odot$) brown dwarfs at all. This figure and caption are comparable to Fig.~9, \citet{Bate2019}.  }
\label{fig:cumIMF}
\end{figure}

\begin{table}
\begin{tabular}{lccccc}\hline
Mass Range ~ [M$_\odot$]& Single & Binary  & Triple & Quadruple  \\ \hline
\multicolumn{5}{c}{Redshift $z=5$, ~Metallicity $Z=0.01~{\rm Z}_\odot$}  \\ \hline
\hspace{0.83cm}$M<0.03$       &      11     &     0     &      0      &     0    \\
$0.03\leq M<0.07$      &    20     &   1    &       0     &      0   \\
$0.07\leq M<0.10$      &      6      &    0      &     0     &      0   \\
$0.10\leq M<0.20$      &      11     &     2    &       1     &      0   \\
$0.20\leq M<0.50$      &      9     &     2    &       1     &     4   \\
$0.50\leq M<0.80$      &     4      &     0      &     3     &      1   \\
$0.80\leq M<1.2$        &       2      &     1      &     1     &      0   \\
\hspace{0.83cm}$M>1.2$        &       2      &    3      &     1    &      2   \\ \hline
\multicolumn{5}{c}{Redshift $z=5$, ~Metallicity $Z=0.1~{\rm Z}_\odot$}  \\ \hline
\hspace{0.83cm}$M<0.03$       &      9     &     0     &      0      &     0    \\
$0.03\leq M<0.07$      &    16     &   0   &       0     &      0   \\
$0.07\leq M<0.10$      &      6      &    0      &     0     &      0   \\
$0.10\leq M<0.20$      &      11     &     2    &      1     &      0   \\
$0.20\leq M<0.50$      &      9     &     0   &       2     &     1   \\
$0.50\leq M<0.80$      &     1      &     1      &     0     &      2   \\
$0.80\leq M<1.2$        &       1     &     1      &     0     &      1   \\
\hspace{0.83cm}$M>1.2$        &       0       &    1      &     6     &      2   \\ \hline
\multicolumn{5}{c}{Redshift $z=5$, ~Metallicity $Z={\rm Z}_\odot$}  \\ \hline
\hspace{0.83cm}$M<0.03$       &     0     &     0     &      0      &     0    \\
$0.03\leq M<0.07$      &    2     &   0    &       0     &      0   \\
$0.07\leq M<0.10$      &     5     &     0    &       0     &      0   \\
$0.10\leq M<0.20$      &      10     &     1    &       0     &     0   \\
$0.20\leq M<0.50$      &      7      &     2      &     1     &      0   \\ 
$0.50\leq M<0.80$      &       2     &     1      &     0     &      0   \\
$0.80\leq M<1.2$        &       3      &     0      &     1     &      0   \\
\hspace{0.83cm}$M>1.2$        &       3       &    2      &     3     &      2   \\ \hline
All masses, 3  calculations                    &   150   &     20     &     21   &      15           \\ \hline
\end{tabular}
\caption{\label{tablemult} The numbers of single and multiple systems for different primary mass ranges at the end of the three $z=5$ radiation hydrodynamical calculations with different metallicities. This table and caption are comparable to Table 2, \citet{Bate2019}.  }
\end{table}

\begin{table}
\begin{center}
\begin{tabular}{lccccl}\hline
Object Number & Mass & $t_{\rm form}$ & Accretion Rate\\
& [M$_\odot$] & [$t_{\rm ff}$] & [M$_\odot$~yr$^{-1}$] \\ \hline
  1 & 2.4253 & 0.8606  & $4.00\times10^{-5}$ \\
  2 & 7.1796 & 0.8616  & $9.83\times10^{-5}$ \\
  3 & 7.1896 & 0.8699  & $9.88\times10^{-5}$ \\
  4 & 2.1061 & 0.8801  & $8.23\times10^{-6}$ \\
  5 & 3.6424 & 0.9140  & $1.28\times10^{-4}$ \\
\hline
\end{tabular}
\end{center}
\caption{\label{tablestars} For each of the three calculations, we provide online tables of the stars and brown dwarfs that were formed, numbered by their order of formation, listing the mass of the object at the end of the calculation, the time (in units of the initial cloud free-fall time) at which it began to form (i.e. when a sink particle was inserted), and the accretion rate of the object at the end of the calculation (precision  $\approx 10^{-7}$~M$_\odot$~yr$^{-1}$).  The first five lines of the table for the $z=5$ solar metallicity calculation are provided above. This table and caption are comparable to Table 3, \citet{Bate2019}.  }
\end{table}

In Fig.~\ref{fig:cumIMF}, we compare the cumulative mass functions at the end of the three calculations, along with the parameterisation of the observed Galactic IMF of \cite{Chabrier2005}.   There is a consistent trend of higher stellar masses being produced at higher metallicity, although there is only a small difference between the mass distributions with $Z=0.01~{\rm Z}_\odot$ and $Z=0.1~{\rm Z}_\odot$.  In the most metal-poor calculation, approximately 1/3 of the objects are brown dwarfs, while in the most metal-rich calculation only 2 out of 67 objects (3\%) are brown dwarfs, and one of these is still accreting when the calculation is stopped (we take brown dwarfs to have masses $M<0.075~{\rm M}_\odot$).  

Performing Kolmogorov-Smirnov tests on each pair of distributions shows that although both the mean and median masses are slightly higher in the $Z=0.1~{\rm Z}_\odot$ calculation than in the $Z=0.01~{\rm Z}_\odot$ calculation (Table \ref{table1}), formally, the two lowest metallicity distributions are consistent with random sampling from a single underlying distribution (probability 38\%).  However, the solar metallicity calculation is significantly different, having probabilities of $4\times 10^{-4}$ and $6\times 10^{-5}$ of being randomly sampled from the same distribution as the $Z=0.1~{\rm Z}_\odot$ and $Z=0.01~{\rm Z}_\odot$ calculations, respectively.  

The two low-metallicity distributions are both in reasonable agreement with the \cite{Chabrier2005} IMF.  Kolmogorov-Smirnov tests that compare the numerical distributions to \citeauthor{Chabrier2005}'s parameterisation give probabilities of 2.6\% ($Z=0.01~{\rm Z}_\odot$) and 20\% ($Z=0.1~{\rm Z}_\odot$) of the numerical distributions being randomly drawn from the \cite{Chabrier2005} IMF.  However, the solar metallicity distribution is inconsistent with being drawn from the \cite{Chabrier2005} IMF, with a Kolmogorov-Smirnov probability of only $1\times 10^{-4}$.  The differences essentially arise from the fact that the median mass of the $Z=0.1~{\rm Z}_\odot$ distribution is in agreement with that of the \cite{Chabrier2005} IMF (0.2~M$_\odot$), while the median mass for the $Z=0.01~{\rm Z}_\odot$ distribution is a little lower and the median of the $Z={\rm Z}_\odot$ distribution is much greater (see Table \ref{table1}).  Of course in these probabilities no account is taken of the observational uncertainty in the Galactic median mass.

Finally, as noted by \cite{Bate2019}, the calculations produce protostellar mass functions (PMFs) rather than IMFs \citep{FleSta1994a,FleSta1994b,McKOff2010} because many of the objects are still accreting when the calculations are stopped and the star formation has not finished (Fig.~\ref{fig:IMF}).  This is particularly true of the solar metallicity calculation.  \cite{Bate2012} showed that the form of the distribution of stellar masses in a calculation similar to those performed here (but not treating the diffuse interstellar medium or having separate gas and dust temperatures) did not change significantly with time during his calculation.  Although the maximum stellar mass and the total number of stars both increased with time, the characteristic (median) mass and the form of the mass function remained similar due to the production of new protostars.  This approximate invariance with time is also true of the calculations of \citet{Bate2019} and the calculations presented here, even in the $z=5$ solar metallicity case.

\begin{figure}
\centering \vspace{-0.2cm}
    \includegraphics[width=8.5cm]{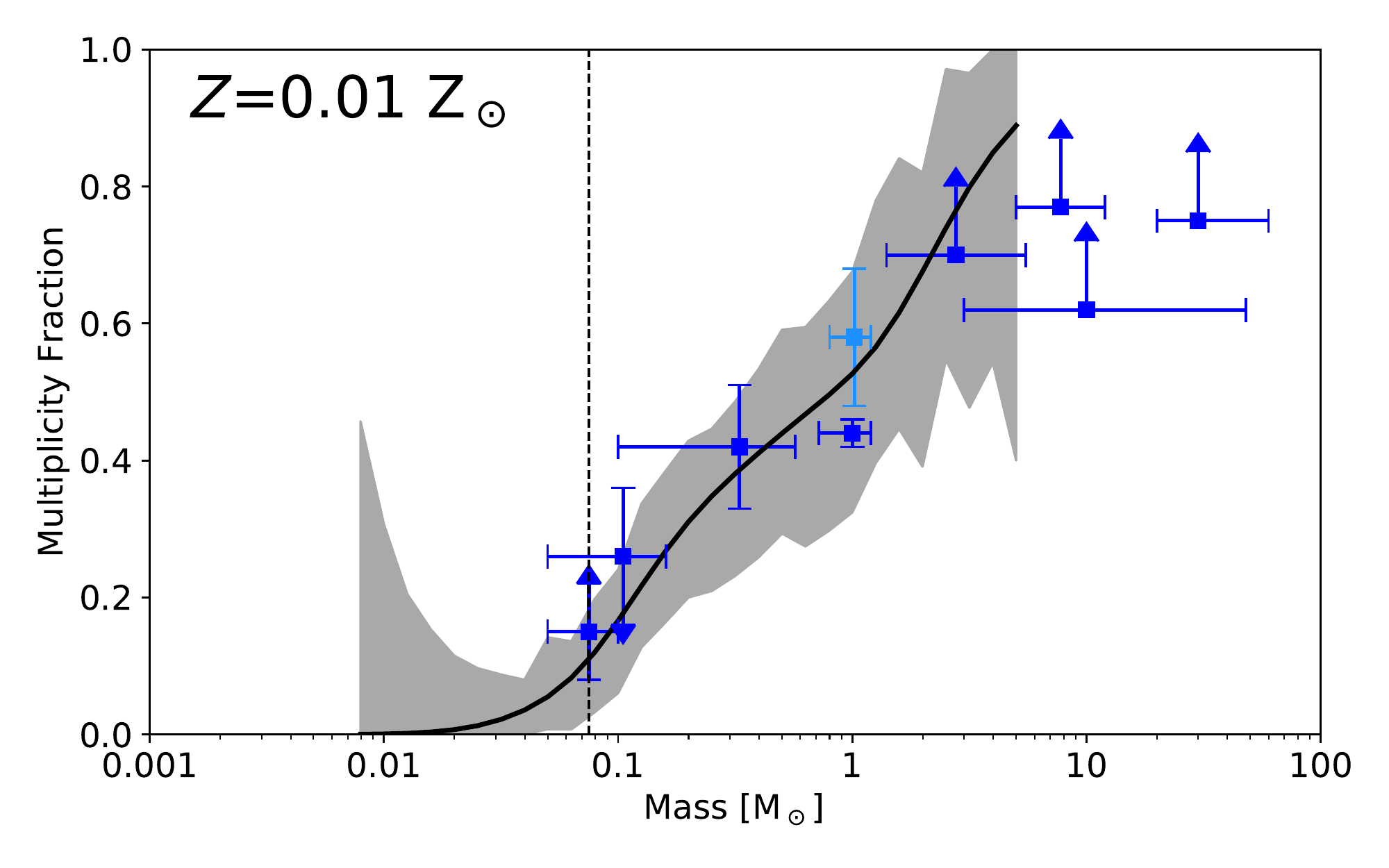}
    \includegraphics[width=8.5cm]{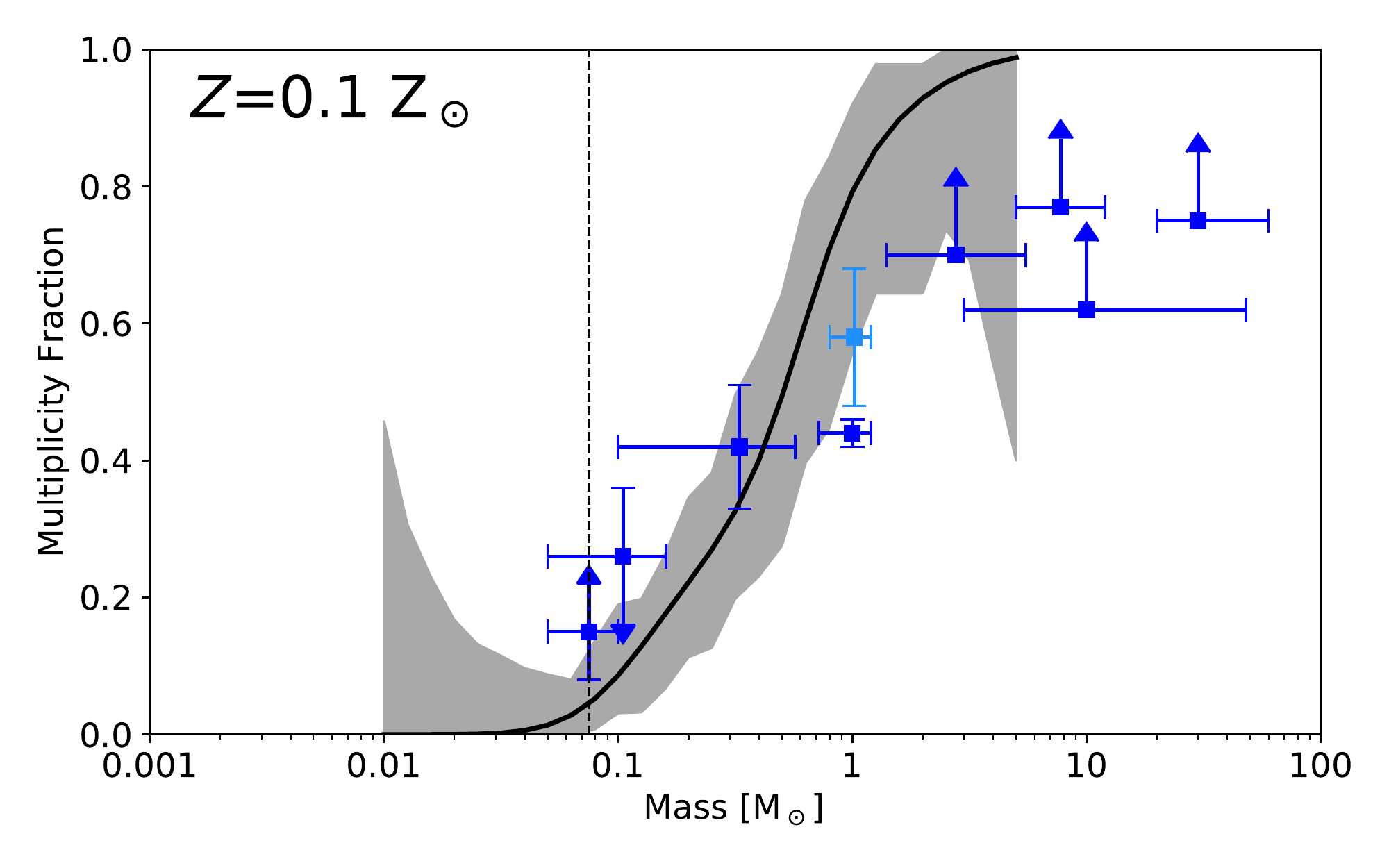}
    \includegraphics[width=8.5cm]{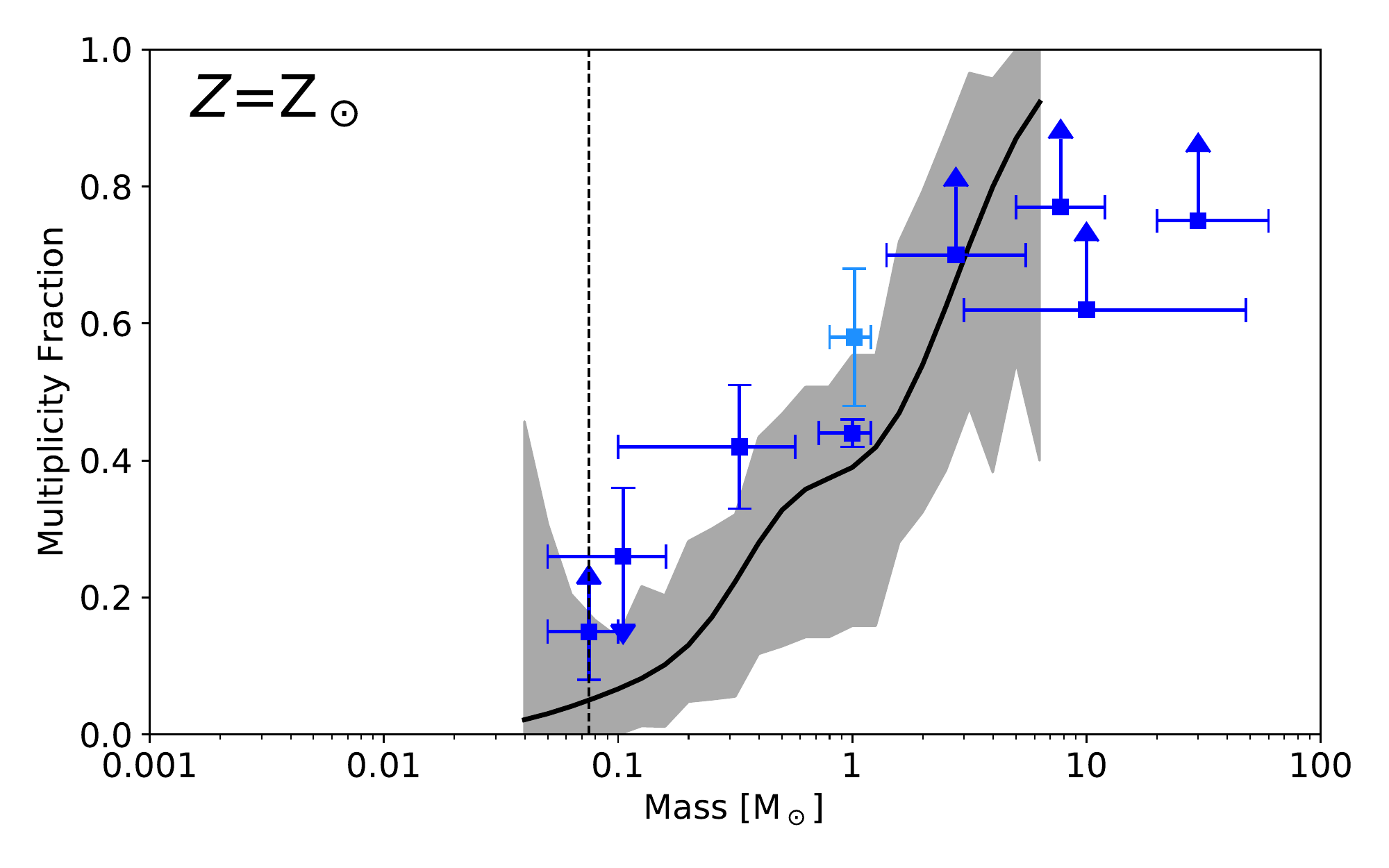}
\caption{Multiplicity fraction as a function of primary mass at the end of each of the calculations with different metallicities.  The thick solid lines give the continuous multiplicity fractions from the calculations computed using a sliding log-normal average and the shaded areas give the approximate $1\sigma$ confidence intervals around the solid line. The blue squares with error bars and/or upper/lower limits give the observed multiplicity fractions from the surveys of \citet{Closeetal2003}, \citet{BasRei2006}, \citet{FisMar1992}, \citet{Raghavanetal2010}, \citet{DuqMay1991}, \citet{Kouwenhovenetal2007}, \citet{Rizzutoetal2013}, \citet{Preibischetal1999} and \citet{Masonetal1998}, from left to right.  Note that the error bars of the \citet{DuqMay1991} results have been plotted using light blue since this survey has been superseded by \citet{Raghavanetal2010}.  The observed trend of increasing multiplicity with primary mass is reproduced by all calculations.  There is no strong difference between the multiplicities obtained with different metallicities, but there seems to be a consistent trend that low-mass stars ($0.1-0.5$~M$_\odot$) have somewhat higher multiplicities with decreasing metallicity. This figure and caption are comparable to Fig.~10, \citet{Bate2019}.  }
\label{multiplicity}
\end{figure}

\begin{table*}
\begin{tabular}{lcccccccccccccl}\hline
Object Numbers & No. of &  No. in & $M_{\rm max}$ & $M_{\rm min}$  & $M_1$ & $M_2$  & $q$ & $a$  & P & $e$  & Relative Spin  & Spin$_1$ & Spin$_2$ \\
& Objects & System & & & & & & & & & or Orbit  & -Orbit & -Orbit \\
&  &  & & & & & & & & & Angle & Angle & Angle\\
     & &    & [M$_\odot$] & [M$_\odot$] & [M$_\odot$] & [M$_\odot$] &  & [au] & [yr] & & [deg] & [deg] & [deg] \\ \hline

 22,  30                     &  2    &  4 &  1.240 &  0.870 &  1.240 &  0.870 &  0.702  &    0.70   &   0.40 &  0.029  &  19  &  63  &  48      \\
 36,  42                     &  2    &  3 &  0.876 &  0.614 &  0.876 &  0.614 &  0.701  &    3.23    &  4.76 &  0.195  &  54  &  59 &   18      \\
( 36,  42),  43           &   3    &  3 &  0.876 &  0.393  & 1.490 &  0.393 &  0.264 &    17.15   &  51.74 &  0.071  &  20 &    -- &   --     \\
( 24,  34), ( 22,  30)   &  4    &  4 &  1.400 &  0.870 &  2.738 &  2.110 &  0.771 &   417.97  &    3880 &  0.576   &  -- &    --  &   --     \\

\hline
\end{tabular}
\caption{\label{tablemultprop} For each of the three calculations, we provide online tables of the properties of the multiple systems at the end of each calculation.  The structure of each system is described using a binary hierarchy.  For each `binary' we give the masses of the most massive star $M_{\rm max}$ in the system, the least massive star $M_{\rm min}$ in the system, the masses of the primary $M_1$ and secondary $M_2$, the mass ratio $q=M_2/M_1$, the semi-major axis $a$, the period $P$, the eccentricity $e$.  For binaries, we also give the relative spin angle, and the angles between orbit and each of the primary's and secondary's spins.  For triples, we give the relative angle between the inner and outer orbital planes. For binaries, $M_{\rm max}=M_1$ and $M_{\rm min}=M_2$.  However, for higher-order systems $M_1$ gives the combined mass of the most massive sub-system (which may be a star, binary, or a triple) and $M_2$ gives the combined mass of the least massive sub-system (which also may be a star, a binary, or a triple).  Multiple systems of the same order are listed in order of increasing semi-major axis.  As examples, we provide selected lines from the table from the $z=5$ solar metallicity calculation. This table and caption are comparable to Table 4, \citet{Bate2019}.  }
\end{table*}

\subsubsection{Multiplicity as a function of primary mass}

The formation mechanisms of multiple systems and the evolution of their properties (e.g. separations) has been discussed in some detail by \cite{Bate2012} and \cite{Bate2018} and will not be repeated here.  Our primary purpose is to determine whether or not variation of the metallicity and/or redshift alters stellar properties significantly.

As in \citet{Bate2009a, Bate2012}, and subsequent papers, to quantify the fraction of stars and brown dwarfs that are in multiple systems, we use the multiplicity fraction, $mf$, defined as a function of stellar mass as
\begin{equation}
mf = \frac{B+T+Q}{S+B+T+Q},
\label{eq:mf}
\end{equation}
where $S$ is the number of single stars within a given mass range and, $B$, $T$, and $Q$ are the numbers of binary, triple, and quadruple systems, respectively, for which the primary has a mass in the same mass range.  As discussed by \cite{HubWhi2005} and \cite{Bate2009a}, this measure of multiplicity is relatively insensitive to both observational incompleteness (e.g., if a binary is found to be a triple it is unchanged) and further dynamical evolution (e.g., if an unstable quadruple system decays the numerator only changes if it decays into two binaries).  We also use the same method for identifying multiple systems as that used by \cite{Bate2009a} and \cite{Bate2012}.  We identify binary, triple, and quadruple stellar systems, but we ignore higher-order multiples (e.g.\ a quintuple system consisting of a triple and a binary orbiting one another is counted as one triple and one binary).  We choose to stop at quadruple systems since it is likely that many higher order systems would be unstable and would eventually decay.

Table \ref{tablemult} provides the numbers of single and multiple star and brown dwarf systems produced by each calculation.  \citet{Bate2019} provided electronic ASCII tables of the properties of each of the stars, brown dwarfs, and multiple systems produced by the calculations discussed in that paper, and we do the same here.  We present tables of the masses, formation times, and final accretion rates of the stars and brown dwarfs (see Table \ref{tablestars} for an example). The tables are given file names such as {\tt Table3\_Stars\_z5\_Metal01.txt} for the $Z=0.1~{\rm Z}_\odot$ calculation.  We also produce tables listing the properties of each multiple system (see Table \ref{tablemultprop} for an example). These tables are given file names such as {\tt Table4\_Multiples\_z5\_Metal1.txt} for the $Z={\rm Z}_\odot$ calculation.

The overall multiplicities for stars and brown dwarfs of all masses from each of the calculations are 26, 27, and 29 per cent for the calculations with metallicities of 1/100, 1/10, 1 times solar, respectively.  The typical $1\sigma$ uncertainties are $\pm 5$ per cent.  Therefore, there is no evidence for a significant dependence of the overall multiplicity on metallicity.  However, because the mass distributions vary between the calculations and multiplicity varies with stellar mass this can be misleading.

In Fig.~\ref{multiplicity}, for each of the three simulations we compare the multiplicity fraction of the stars and brown dwarfs as functions of the stellar mass of the primary with the values obtained from various Galactic observational surveys (see the figure caption).  The results from each of the calculations have been plotted using a thick solid line that gives the continuous multiplicity fraction computed using a sliding log-normal-weighted average.  The width of the log-normal average is half a decade in stellar mass.  The shaded region gives the approximate $1\sigma$ (68\%) uncertainty on the sliding log-normal average.

\begin{figure*}
\centering
   \includegraphics[width=5.8cm]{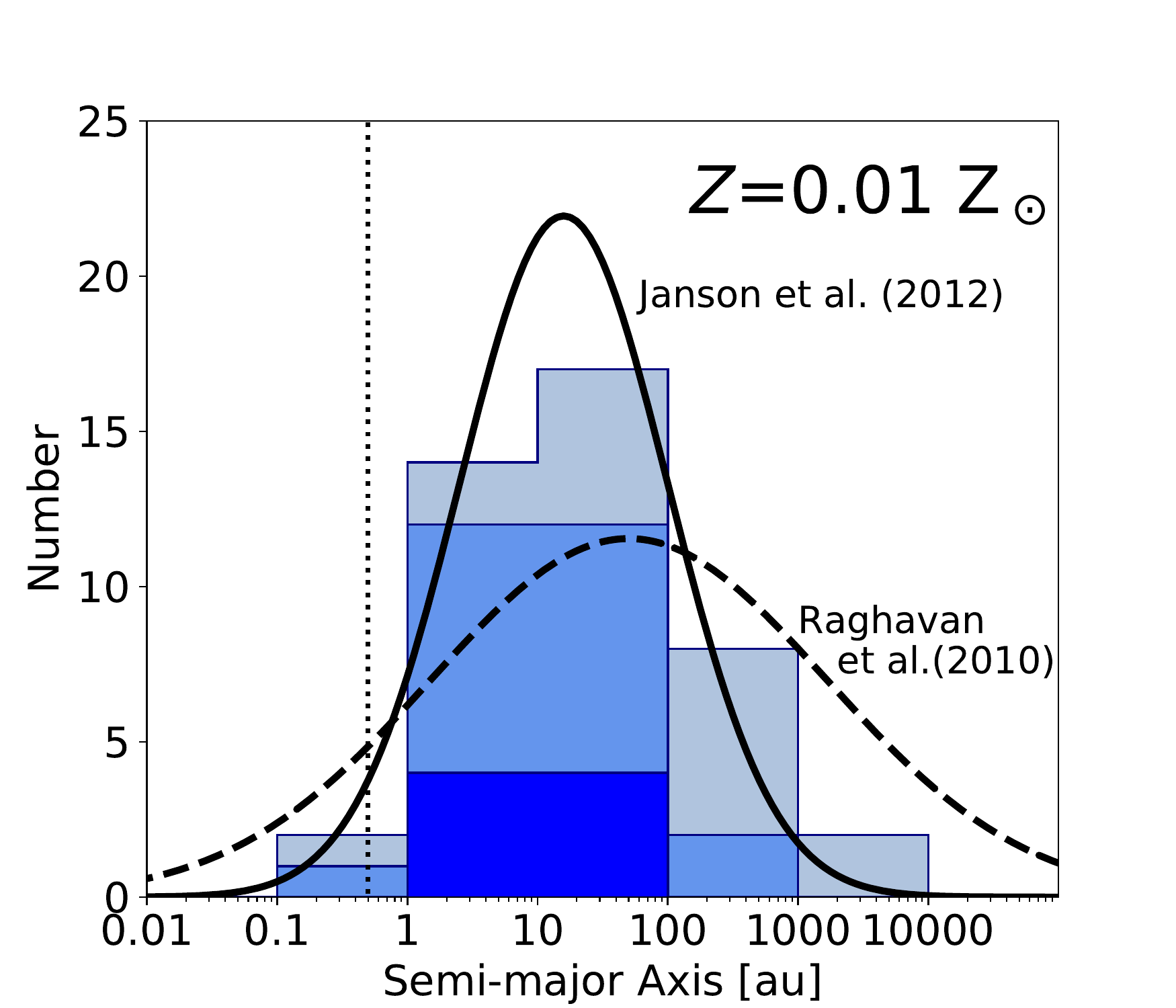} 
   \includegraphics[width=5.8cm]{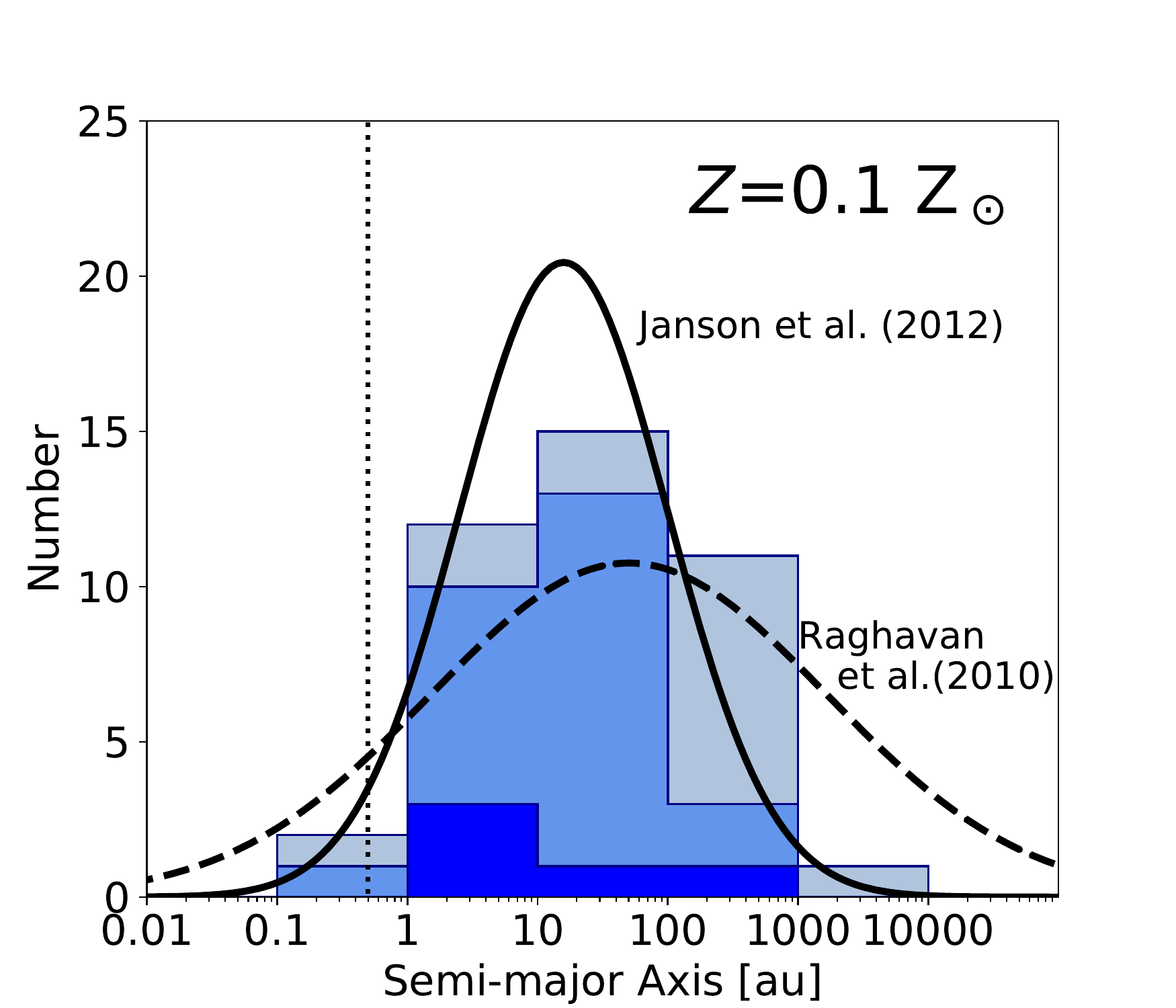} 
   \includegraphics[width=5.8cm]{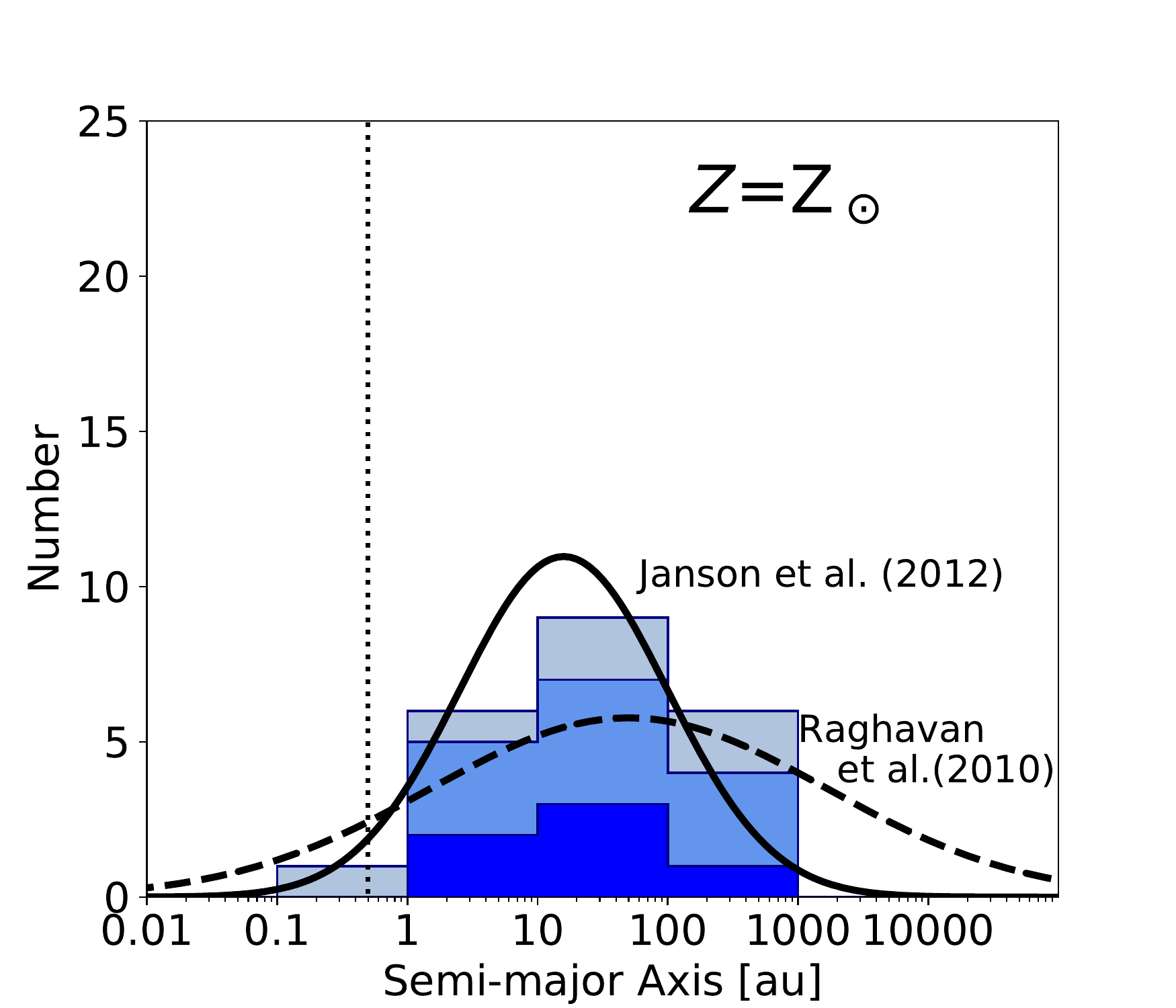} 
\caption{The distributions of separations (semi-major axes) of multiple systems with stellar primaries ($M_*>0.1$~M$_\odot$) produced by the calculations with different metallicities.  The dark, medium, and light histograms give the orbital separations of binaries, triples, and quadruples, respectively (each triple contributes two separations, each quadruple contributes three separations).  The solid curve gives the M-dwarf separation distribution (scaled to match the same area as each of the histograms) from the M-dwarf survey of \citet{Jansonetal2012}, and the dashed curve gives the separation distribution for solar-type primaries of \citet{Raghavanetal2010}. 
Note that since most of the simulated systems are low-mass, the distributions are expected to match the \citeauthor{Jansonetal2012} distribution better than that of \citeauthor{Raghavanetal2010}  The vertical dotted line gives the resolution limit of the calculations as set by the accretion radii of the sink particles (0.5 au). This figure and caption are comparable to Fig.~11, \citet{Bate2019}.  }
\label{separation_dist}
\end{figure*}

All of the calculations produce multiplicity fractions that increase strongly with increasing primary mass, in qualitative agreement with observed stellar systems.  The actual values of the multiplicities are in good agreement with the observed multiplicities of Galactic field stars for the lowest metallicity calculation, and they are also similar to the multiplicities obtained for all of the present-day star calculations of \citet{Bate2019} with metallicities ranging from $Z=0.01 - 3$~Z$_\odot$.  However, unlike the present-day star formation calculations of \citet{Bate2019}, the multiplicities of the three calculations at redshift $z=5$ do seem to depend on metallicity.  There is no consistent trend of the multiplicities for solar-mass and more massive stars, but for primaries with masses $M_1 \lsim 0.1-0.5$~M$_\odot$ (i.e. M-dwarfs) the multiplicities consistently increase with decreasing metallicity such that the multiplicities at metallicity $Z=0.01$~Z$_\odot$ are approximately twice as high as for $Z={\rm Z}_\odot$.

Only a single very-low-mass (VLM; $M_1 < 0.1~{\rm M}_\odot$) multiple system is produced across all three simulations (see Table \ref{tablemult}), and this is in the lowest metallicity calculation.  Fig.~\ref{multiplicity} shows that the VLM multiplicity for metal-poor ($Z=0.01~{\rm Z}_\odot$) star formation at redshift $z=5$ is similar to the Galactic value.  However, at higher metallicities we find that the VLM multiplicity produced by star formation at redshift $z=5$ is lower than both Galactic field value and that obtained from the present-day star formation calculations of \citet{Bate2019}.

\begin{figure}
\centering
    \includegraphics[width=8.5cm]{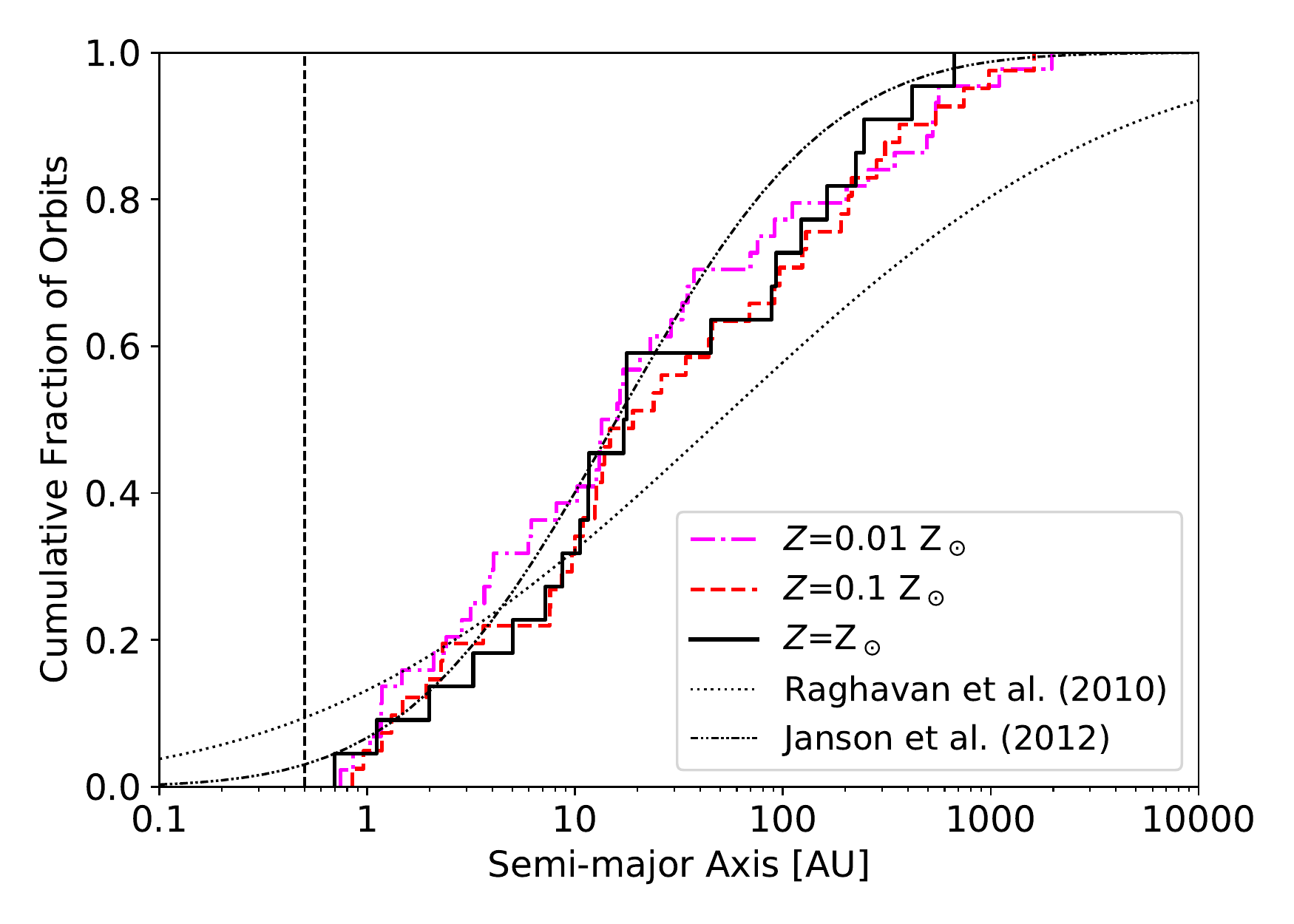} \vspace{-0.5cm}
\caption{The cumulative semi-major axis (separation) distributions of the multiple systems produced by the three $z=5$ calculations with different metallicities.  All orbits are included in the plot (i.e. two separations for triple systems, and three separations for quadruple systems), and the lines for different metallicities include both stars and brown dwarfs.  The vertical dashed line marks the resolution limit of the calculations as determined by the accretion radii of the sink particles.  Performing Kolmogorov-Smirnov tests on pairs of distributions shows that the distributions for the different metallicities are statistically indistinguishable. This figure and caption are comparable to Fig.~12, \citet{Bate2019}.  }
\label{cumsep_comp}
\end{figure}

\subsubsection{Separation distributions of multiple systems}
\label{sec:separations}

In Fig.~\ref{separation_dist}, we present the separation (semi-major axis) distributions of the stellar ($M_1> 0.1$~M$_\odot$) multiples (there is only one VLM multiple, the binary in the $Z=0.01~{\rm Z}_\odot$ calculation).  The distributions are compared with the log-normal distributions from the surveys of M-dwarfs by \cite{Jansonetal2012} and solar-type stars by \cite{Raghavanetal2010}.  Although binaries as close as 0.03~au (6~R$_\odot$) can be modelled without being merged, gas is not modelled within the sink particle accretion radii of 0.5~au.  The lack of gas to provide dissipation on small scales inhibits the formation of very close systems \citep*{BatBonBro2002b}.  

There is no obvious dependence of the shape of the distributions on metallicity (in all cases the peak occurs in the $10-100$~au range of semi-major axis).  The absolute numbers are lower with increasing metallicity due to the smaller number of stars produced at higher metallicity (Table \ref{table1}).  The peak occurs at a similar separation to the peak in the Galactic population of stellar multiples.  The spread appears slightly wider than for Galactic M-dwarf systems and narrower than Galactic solar-type stars, consistent with a mixture of primary stellar masses that is dominated by low-mass stars.

In Fig.~\ref{cumsep_comp}, we provide the cumulative separation distributions for each of the calculations (this time including the VLM binary from the $Z=0.01~{\rm Z}_\odot$ calculation).  Kolmogorov-Smirnov tests performed on each pair of distributions shows that they are statistically indistinguishable.  They are also each in good agreement with the separation distribution of M-dwarfs as determined by \citet{Jansonetal2012}, and they tend to have slightly closer separations than the solar-type stars \citep{Raghavanetal2010}.  In the $Z=0.01~{\rm Z}_\odot$ calculation, most of the multiple systems have primary masses $M_1=0.1-0.8~{\rm M}_\odot$, while in each of the other two calculations roughly half of the multiples are low-mass and half have primary masses $M_1>0.8~{\rm M}_\odot$. 

For present-day star formation, \citet{Bate2019} found a weak possible trend for multiple systems to be preferentially tighter at lower metallicities, though the small numbers of systems limited the confidence in the result.  There is no sign of such a trend for the systems formed at redshift $z=5$. 

\begin{figure}
\centering
    \includegraphics[width=8.5cm]{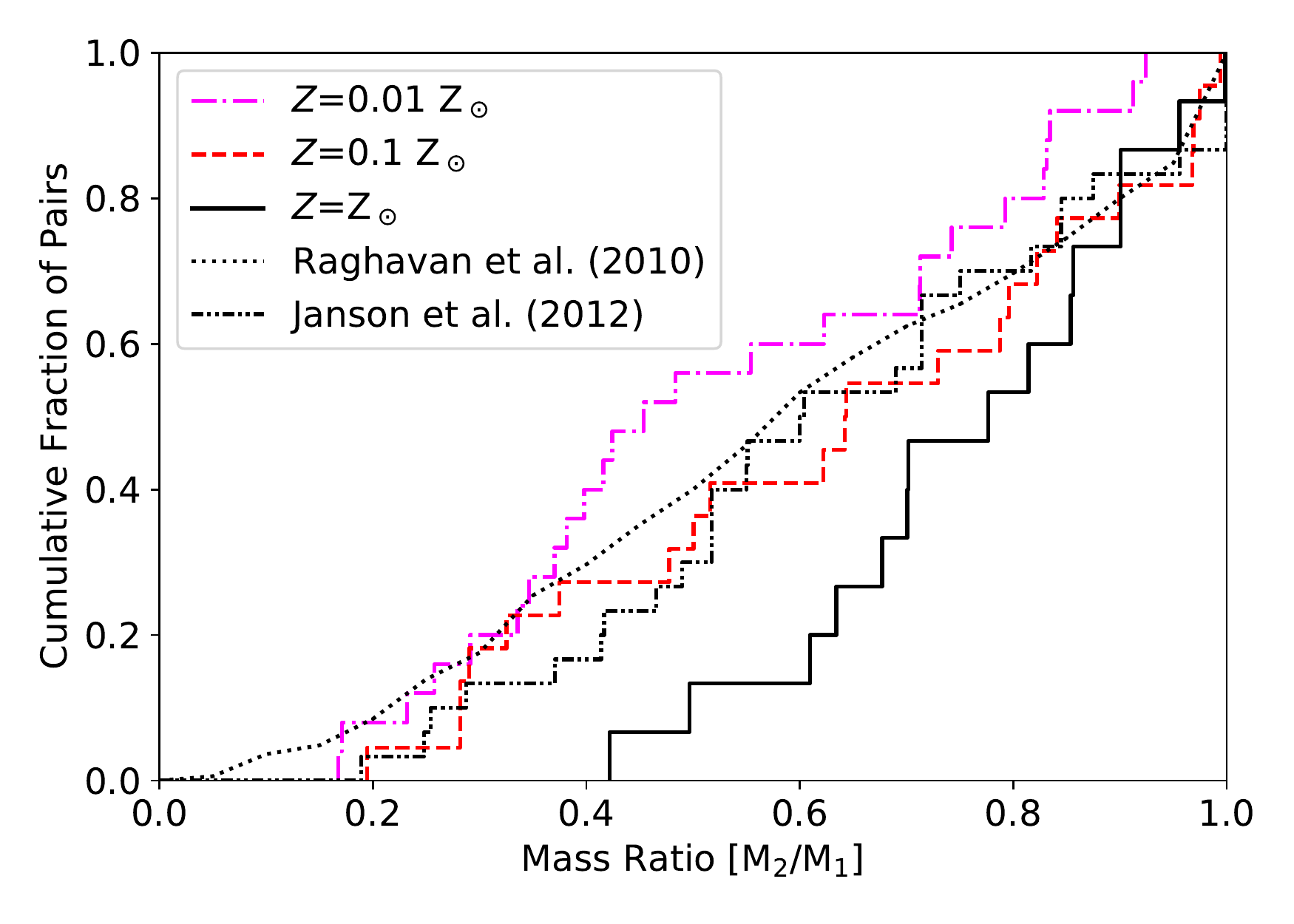} \vspace{-0.5cm}
\caption{The cumulative mass ratio distributions of pairs of objects produced by the three $z=5$ calculations with different metallicities.  Pairs include both binaries and bound pairs that are components of triple or quadruple systems.  We also plot the observed mass ratio distribution of solar-type stars from \citet{Raghavanetal2010}. There is a clear and consistent trend for the more metal-rich systems to have pairs consisting of more equal-mass components.  However, performing Kolmogorov-Smirnov tests on each pair of the simulated distributions, only the $Z=0.01~{\rm Z}_\odot$ and $Z={\rm Z}_\odot$ distributions are inconsistent with being drawn from the same underlying distribution (Kolmogorov- Smirnov probability of 1.3 percent) due to the relatively small numbers of pairs that are produced by the calculations. This figure and caption are comparable to Fig.~14, \citet{Bate2019}.   }
\label{cumq}
\end{figure}

\subsubsection{Mass ratio distributions of pairs}
\label{sec:q}

In Fig.~\ref{cumq} we plot the cumulative mass ratio distributions for each calculation (including all primary masses, and including the single VLM binary from the $Z=0.01~{\rm Z}_\odot$ calculation).  These distributions include binary systems, and pairs that are the inner components of triple and quadruple systems.  A triple system composed of a binary with a wider companion contributes the mass ratio from the closest pair, as does a quadruple composed of a triple with a wider companion. A quadruple composed of two pairs orbiting each other contributes two mass ratios -- one from each of the pairs.  We also plot the cumulative observed mass-ratio distributions for M-dwarfs from \cite{Jansonetal2012}, and for solar-type primaries from \cite{Raghavanetal2010}.

There is a clear progression in the mass ratio distributions from the $z=5$ calculations such that the more metal-rich calculations tend to have pairs consisting of more equal-mass components.  The mass ratio distribution of the $Z=0.1~{\rm Z}_\odot$ is most similar to the mass ratio distribution of M-dwarfs from \cite{Jansonetal2012}.  The $Z=0.01~{\rm Z}_\odot$ calculation has a greater fraction of $q<0.5$ pairs, while the $Z={\rm Z}_\odot$ calculation has almost no pairs with $q<0.5$.  Despite the clear progression with metallicity, formally only the $Z=0.01~{\rm Z}_\odot$ and $Z={\rm Z}_\odot$ distributions are inconsistent with being drawn from the same underlying distribution (Kolmogorov-Smirnov probability of 1.3 percent) due to the relatively small numbers of pairs that are produced by the calculations.

For present-day  ($z=0$) star formation, \citet{Bate2019} found that all of the mass ratio distributions from the four calculations ranging in metallicity from $Z=0.01 - 3 ~ {\rm Z}_\odot$ were statistically indistinguishable, although again the lowest metallicity calculation had the greatest fraction of low mass-ratio systems.  All of the distributions were formally consistent with being drawn from the \cite{Raghavanetal2010} distribution, except for the highest metallicity calculation ($Z=3~{\rm Z}_\odot$) which had a deficit of low mass ratio pairs. Thus, the general trends found (a greater fraction of low mass ratio pairs in lower metallicity calculations) are the same for both present-day star formation and for redshift $z=5$.  We have also performed Kolmogorov-Smirnov tests on the mass ratio distributions for the $z=0$ and $z=5$ distributions that were obtained using the same metallicity (i.e. $Z=0.01,0.1, 1~{\rm Z}_\odot$).  In each case the mass ratio distributions are indistinguishable at the two redshifts, although again the sensitivity of this test is limited by the relatively small numbers of pairs that are produced by the $z=5$ simulations.

\begin{figure}
\centering
    \includegraphics[width=8.6cm]{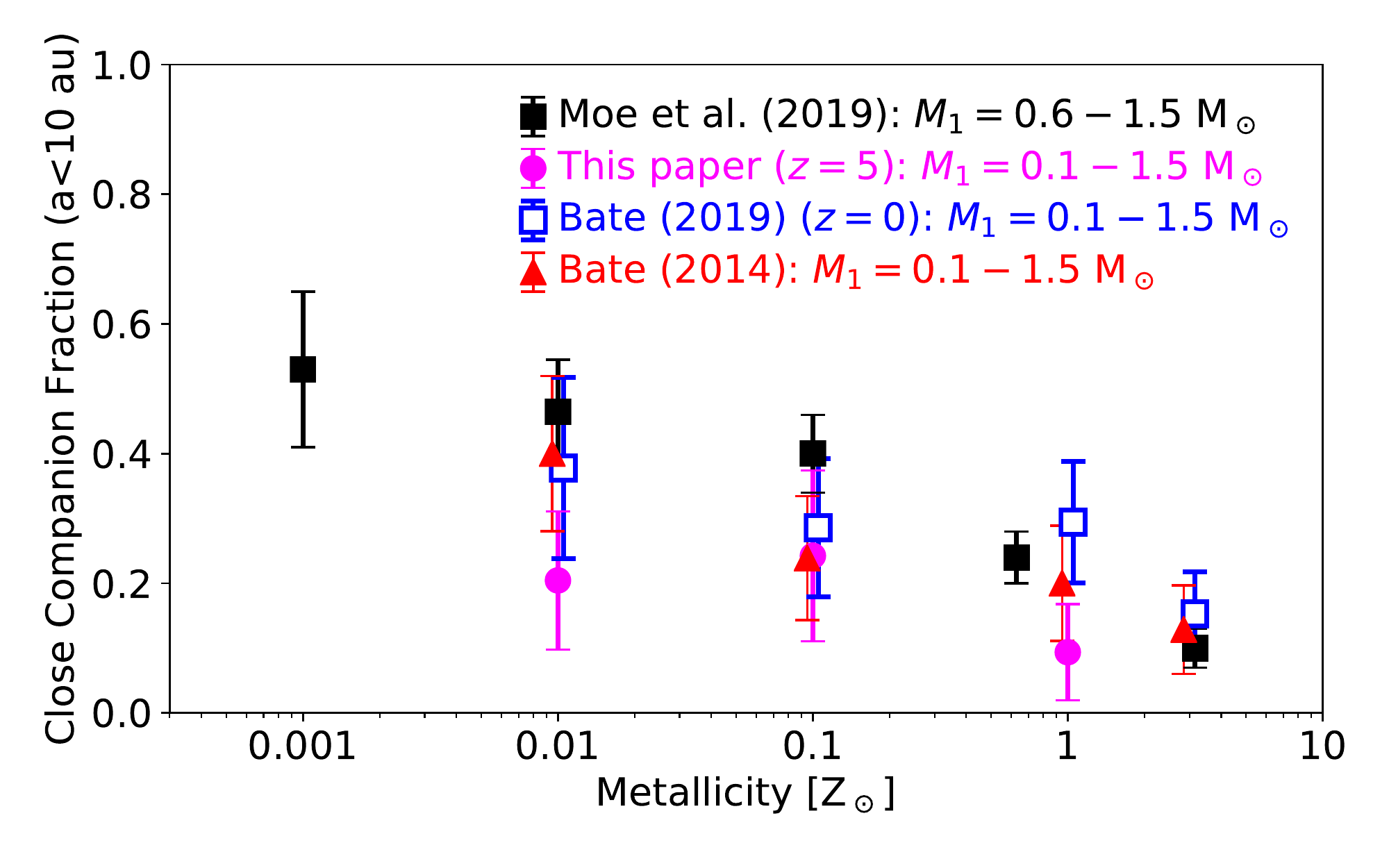} \vspace{-0.5cm}
\caption{ The frequencies of close companions (semi-major axes $a<10$~au) for low-mass stars with masses $M_*=0.1-1.5$~M$_\odot$ that are produced by the three $z=5$ calculations with different metallicities (magenta filled circles and errorbars).  We compare the results with the equivalent frequencies from the numerical simulations of present-day ($z=0$) star formation by \citet{Bate2019} (blue open squares with errorbars) and \citet{Bate2014} (red triangles and dashed errorbars), and with the observed values for stellar masses $M_*=0.6-1.5$~M$_\odot$ from \citet{MoeKraBad2019} (black filled squares and errorbars). The error bars for the numerical simulations give 95 percent confidence intervals.  The values of the metallicity have been slightly offset horizontally for the \citet{Bate2014,Bate2019} results for clarity.   The results from the simulations are consistent with the observed anti-correlation between the frequency of close companions and metallicity, and the numerical values from the present-day calculations are in reasonable agreement with the observed values.  For the redshift $z=5$ calculations, the frequencies are on average approximately a factor of two lower and the anti-correlation is less clear. This figure and caption are comparable to Fig.~15, \citet{Bate2019}.  }
\label{fig:closebinaries}
\end{figure}

\subsubsection{Close binaries}
\label{sec:closebinaries}

\begin{figure*}
\centering\vspace{-0.3cm}
   \includegraphics[width=8cm]{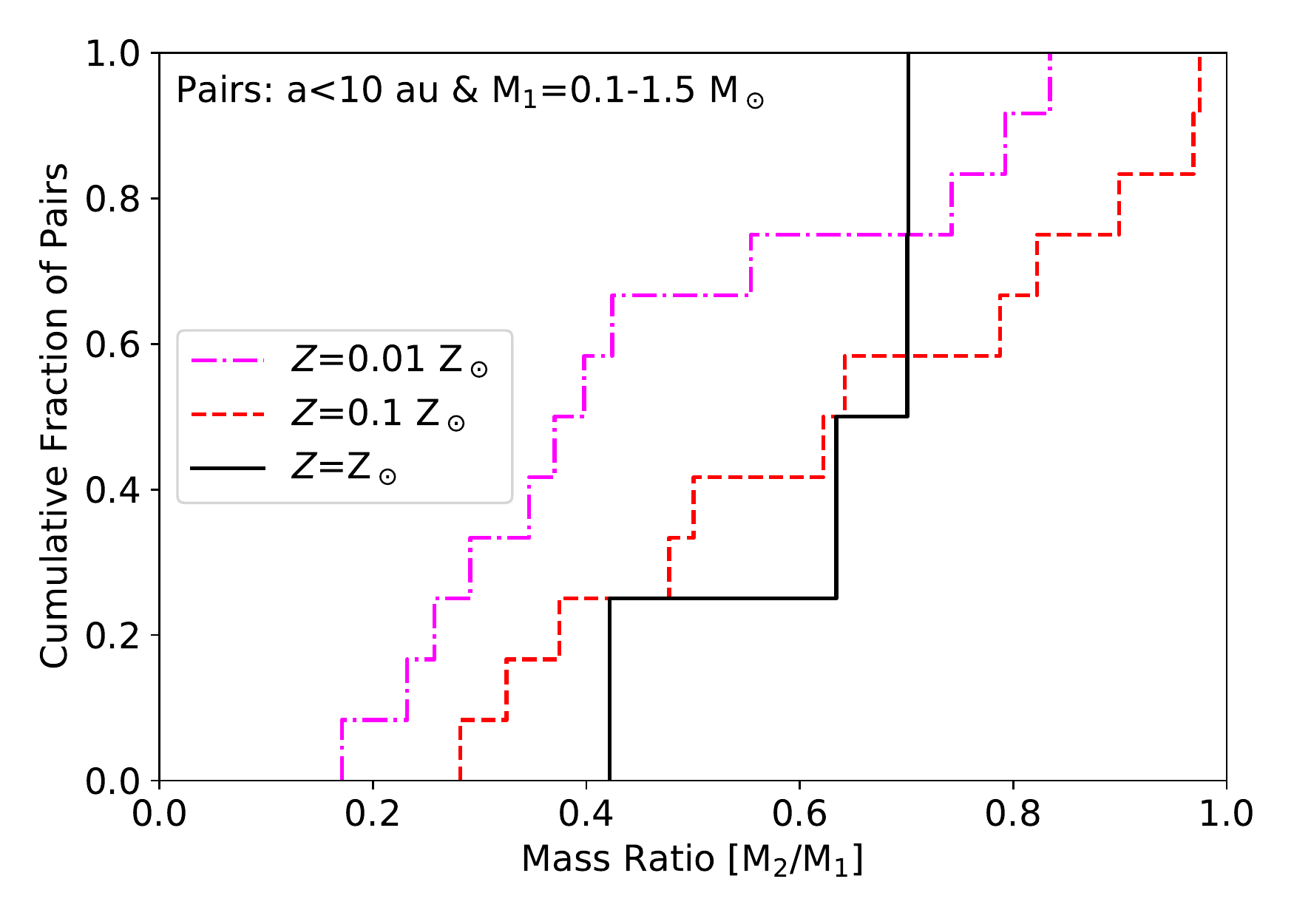} \hspace{0.5cm}
    \includegraphics[width=8cm]{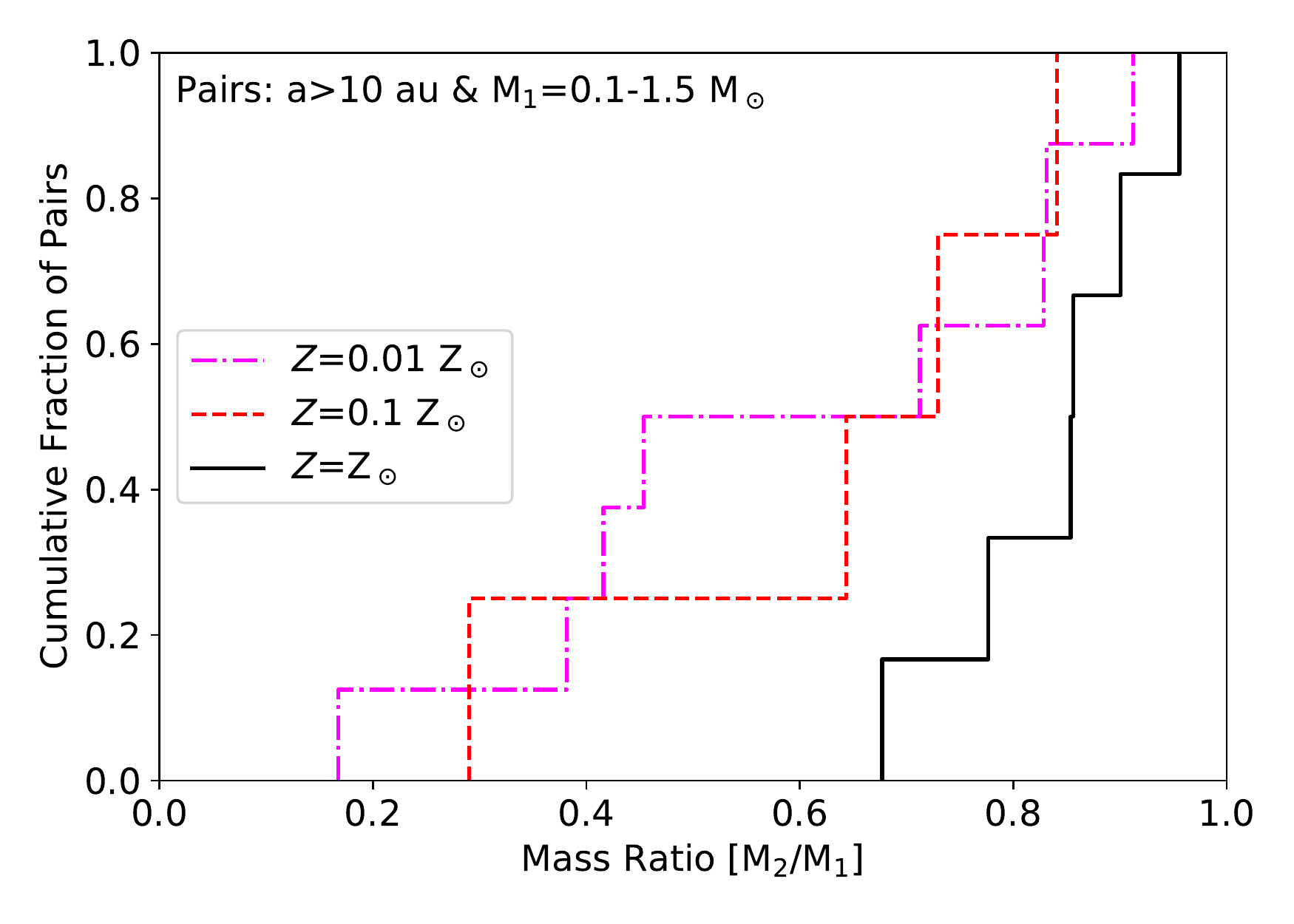}
    \vspace{-0.3cm}
\caption{ The mass ratio distributions of stellar pairs with primary masses $M_1=0.1-1.5$~M$_\odot$ that are produced by the four calculations with different metallicities.  In the left panel, we give the distributions for pairs with semi-major axes $a<10$~au, while in the right panel we give the distributions for pairs with $a>10$~au. Due to the small numbers of objects, the probability of any two pairs of distributions being drawn from the same underlying distribution never falls below 9 percent in the left plot.  However, the close pairs (left panel) do display a consistent trend of more low mass ratio systems as the metallicity is reduced, whereas there is no clear trend for the wider pairs (right panel). This figure and caption are comparable to Fig.~16, \citet{Bate2019}.  }
\label{fig:CBq}
\end{figure*}

\cite{Bate2019} found that although the stellar multiplicities for present-day star formation with metallicities ranging from $Z=0.01-3~{\rm Z}_\odot$ were indistinguishable when considering all separations, if the frequency of close pairs with semi-major axes less than 10~au were considered then the binary frequency was anti-correlated with metallicity.  \cite{Bate2019} investigated this particular separation range because three observational papers that had been published during the preceding year had found that the close binary fractions for solar-type stars were anti-correlated with metallicity \citep{Badenes_etal2018, ElBRix2019,MoeKraBad2019}.  This type of anti-correlation had been claimed before \citep{GreLin2007,Raghavanetal2010}, but the small numbers of stars involved and potential observational biases had limited the confidence in the results.

In addition, \cite{Bate2019} found that close pairs (semi-major axes, $a<10$~au) displayed a consistent trend such that there were more low mass-ratio systems for lower metallicities, but that there was no such trend for wider pairs.

In the redshift $z=5$ calculations analysed in this paper, we have already seen that the multiplicities of low-mass stars (M-dwarfs) appear to be lower at higher metallicity (Section \ref{multiplicity}) even considering all separation ranges.  Here we examine just the close pairs ($a<10$~au).   We also restrict our study to systems with primary with masses $M_1\approx 0.1-1.5$~M$_\odot$, which is larger than the range of solar-type primaries, $M_1\approx 0.6-1.5$~M$_\odot$, that was considered by \cite{MoeKraBad2019}.  We use a larger range because we have a comparatively small number of systems (this larger range was also used by \citealt{Bate2019}).

In Fig.~\ref{fig:closebinaries}, we present the frequencies of close pairs (semi-major axes $<10$~au) in systems with primary masses $M_1 = 0.1-1.5$~M$_\odot$ for each of our simulations.  We compare these to the fractions that were presented in the equivalent figure by \cite{Bate2019} for the the present-day star formation simulations analysed by that paper, the earlier calculations of  \cite{Bate2014}, and also to the observational results of \cite{MoeKraBad2019}.  The results from the $z=5$ simulations are consistent with an anti-correlation between close binary frequency and metallicity, but the results are less convincing than for the \cite{Bate2014,Bate2019} simulations.  The close companion fractions are generally lower for the $z=5$ calculations than for the present-day star formation simulations.  In other words, star formation at $z=5$ results in a lower frequency close pairs than present-day star formation by about a factor of two for the same metallicity.  On top of this, there is evidence for an anti-correlation with metallicity since the frequency in the solar-metallicity calculation is about 1/3 of that obtained from the $Z=0.1~{\rm Z}_\odot$.  However, the frequencies are similar for the two lowest metallicity calculations.  Overall, the results are consistent with an anti-correlation between close binary frequency and metallicity for star formation at both redshift $z=0$ and $z=5$, but larger numbers of systems are required to be certain.

Following \cite{Bate2019}, in Fig~\ref{fig:CBq}, we give the mass ratio distributions of pairs with low-mass stellar primaries ($M_1=0.1-1.5$~M$_\odot$) for close pairs (semi-major axes $a<10$~au; left panel) and wider pairs (right panel).  For the close pairs, the lowest metallicity calculation produces more low-mass ratio systems than the intermediate metallicity calculation, whereas there is no clear difference for the wider systems.  This is the same trend that was found by \citet{Bate2019} for present-day star formation.  But performing Kolmogorov-Smirnov tests on all pairs of cumulative distributions for the $z=5$ calculations shows that, in each case, the two distributions are consistent with being drawn from the same underlying distribution due to the small numbers.

As discussed by \cite{Bate2019}, the fact that anti-correlation between close pair frequency and metallicity was found in both the \cite{Bate2014} and \cite{Bate2019} sets of calculations implies that the physical mechanism originates from the metallicity dependence of the opacity used in the radiative transfer (which applied in both sets of calculations), rather than from the separate treatment of gas and dust temperatures or the model of the diffuse interstellar medium that was employed in the calculation of \cite{Bate2019} and in this paper.  \citet{Bate2019} found that two different effects of the dust opacities combined to produce the anti-correlation.  First, at higher metallicities the higher dust opacities gave less small-scale fragmentation because dense gas is more optically thick and less able to cool quickly.  Second, at higher metallicities the higher dust opacities result in slower cooling rates of first hydrostatic cores (FHSCs) and, thus, longer lifetimes.  Since FHSCs typically have radii of $\approx 5$~au \citep{Larson1969}, FHSCs need to undergo collapse to form stellar cores before close binaries ($a<10$~au) can be formed.  Longer lifetimes mean that even if small scale fragmentation produces multiple FHSCs, those with high-metallicity have longer to merge into a single FHSC, thereby producing a single protostar rather than a close binary.  Conversely, low-metallicity FHSCs can quickly collapse to produce stellar cores, potentially avoiding merging with another nearby FHSC and producing a close pair.  It is also worth noting that \citet{Bate2019} found that most close pairs were formed by cloud or filament fragmentation, not by disc fragmentation.

In the redshift $z=5$ calculations, there is still evidence for an anti-correlation between close binary frequency and metallicity, but it may be weaker than for present-day star formation and the close binary frequency may be lower (by approximately a factor of two) at a given metallicity at $z=5$ than at $z=0$.  This may be because the first of the two mechanisms leading to the anti-correlation is weakened at higher redshift.  In other words, the slower cooling rates of the FHSCs at higher metallicity leading to more FHSC merges at high metallicity would still operate (giving fewer close pairs at higher metallicity).  But with the higher CMBR temperature there is less small-scale fragmentation of high-density gas because the dense gas is less able to cool, particularly at high metallicity.

\section{Discussion}
\label{sec:discussion}

\cite{Bate2019} performed radiation hydrodynamical calculations of present-day star formation in molecular clouds whose metallicity was varied from $Z=0.01-3~{\rm Z}_\odot$ and found that the statistical properties of the stars were largely independent of the metallicity \citep[see also][]{Myersetal2011, Bate2014}.  The resulting stellar mass distributions, overall stellar multiplicity, and mass ratio distributions of bound pairs were statistically indistinguishable.  Furthermore, the stellar mass distributions were all very similar to the parameterisation of the Galactic IMF of \citet{Chabrier2005}.  The only clear differences that were found was an anti-correlation between the frequency of close ($a<10$~au) protostellar pairs and metallicity, in qualitative agreement with observations of Galactic field stars, a preference for a greater fraction of unequal-mass close pairs with decreasing metallicity, and an increase in the number of protostellar mergers per star with decreasing metallicity.

By contrast, in this paper we have presented results from simulations that are identical to those of \citet{Bate2019} except that the cosmic microwave background radiation was hotter because they were performed at a redshift of $z=5$.  With the hotter CMBR we find that the statistical properties of the stars do vary with metallicity.  In particular, at $Z=0.01~{\rm Z}_\odot$ the stellar mass distributions obtained at redshifts $z=0$ and $z=5$ are indistinguishable (and both are similar to the observed Galactic IMF), but at $z=5$ increasing the metallicity results in an increase in the characteristic stellar mass to the point that brown dwarfs are very rare at solar metallicities.  Furthermore, the multiplicities of M-dwarfs are approximately a factor of two lower at solar metallicity than at $Z=0.01~{\rm Z}_\odot$.  In agreement with the present-day star formation calculations, the mass ratios of stellar pairs are more unequal and there are more stellar mergers per star at lower metallicity.  The anti-correlation between the frequencies of close pairs and metallicity still seems to be present, but it is less clear than in the $z=0$ calculations.

Perhaps the most surprising result from the two papers is that the different level of CMBR {\it qualitatively changes} the dependence of the stellar mass function on metallicity:  the characteristic stellar mass is predicted to be independent of metallicity at $z=0$, but increase with metallicity at $z=5$.  Or, alternately, at low metallicity ($Z=0.01~{\rm Z}_\odot$) the stellar mass function doesn't vary with redshift, but at solar-metallicity the characteristic stellar mass increases with increasing redshift due to the hotter CMBR.

In the following sections, we examine the predictions of past theoretical studies for the dependence of the IMF on redshift, we discuss the implications of these results for how the IMF may vary continuously with redshift and metallicity, and we examine the observational evidence for variations in the IMF and how these results may aid in their interpretation. 

\subsection{Comparison with previous theoretical models}

It is common in the literature to postulate that the characteristic stellar mass is proportional to the typical Jeans mass in a molecular cloud.  The Jeans mass depends on temperature: for example, it can be expressed as varying with temperature and density as $\propto T^{3/2} \rho^{-1/2}$, or with temperature and pressure as $\propto T^2 P^{-1/2}$, or with temperature and surface density as $\propto T^2\mu^{-1}$ \citep{Larson1998}.  Given the strong dependence of the Jeans mass on temperature, if the characteristic stellar mass scales with the Jeans mass it is natural to postulate that the characteristic mass of the IMF should increase with redshift because the cosmic microwave background radiation scales with redshift as $(1+z)$ (equation \ref{eq:TCMBR}) and this exceeds the lowest temperatures found in Galactic molecular clouds ($\sim 6-8$)~K by redshift $z=2-3$.  Similarly, because metal-poor gas is less able to cool than metal-rich gas when it is optically thin, one can make the argument that metal-poor gas is generally expected to be hotter and, therefore, the characteristic stellar mass would also increase with decreasing metallicity.  Indeed, both these arguments are made by \citet{Larson1998} when he discusses the possibility of a time-varying IMF, and assuming that metallicity generally decreases with increasing redshift both of these affects act together to suggest that the characteristic stellar mass should increase with increasing redshift (see also \citealt{Baugh_etal2005}; \citealt*{NarDav2012}).

However, as we have seen from the results in this paper the dependence of the characteristic stellar mass is apparently more complicated than this.  For one thing, although metal-poor gas is a poor at cooling compared metal-rich gas at low densities, at high densities when the cooling of the gas becomes dominated by collisions with dust grains, metal-poor gas remains optically thin to higher densities and, therefore, is able to cool more quickly, leading to lower gas temperatures and, potentially, more fragmentation than metal-rich gas.  \citet{Bate2014} demonstrated this by performing simple spherically-symmetric calculations of collapsing \mbox{1-M$_\odot$} Bonnor-Ebert spheres with radiative transfer and varying opacities, showing that gas temperatures reduce by factors of 5--7 at densities ranging from $10^{-13}$ to $10^{-9}$~g~cm$^{-3}$ when the opacity is reduced from 3 to 0.01 times the opacity of solar-metallicity gas \citep[see Fig. 23 of][]{Bate2014}. 

These opposing effects of metallicity (i.e.\ typically hotter low-density gas, but cooler high-density gas) leads to a more complicated dependence of the characteristic mass on redshift and metallicity than that postulated by \citet{Larson1998}.

Very recently, \citet{ShaKru2022} have investigated analytically how the characteristic stellar mass may vary with metallicity, decreasing from a comparatively high value at low metallicity ($Z \lsim 10^{-4}~{\rm Z}_\odot$) to a much lower value once dust dominates the cooling ($Z \gsim 10^{-2}~{\rm Z}_\odot$).  They argue that this may help to explain the bottom-heavy IMFs apparently observed in massive elliptical galaxies (see Section \ref{sec:galactic} below).  However, they do not consider at all how a change in redshift (i.e. the level of CMBR) may also alter the characteristic mass and its metallicity dependence.  The results from this paper show that both redshift and metallicity need to be taken into account.

In another recent paper, \citet{Guszejnov_etal2022} studied how environment, initial conditions, and feedback may alter the stellar IMF.  Among other results, they find that the characteristic stellar mass varies with metallicity and with the level of the ISRF.  They find that with a typical Galactic ISRF, lower metallicity tends to raise the gas temperature and this in turn produces an increase characteristic stellar mass.  This result is at odds with the results of \cite{Bate2019} who find that under present-day conditions the characteristic stellar mass is quite independent of the metallicity (ranging from 1/100 to 3 times the solar value).  This may be due to the limited resolution of the  \citet{Guszejnov_etal2022} study -- as mentioned above, \cite{Bate2019} found that at low metallicity the effects of intermediate-density gas being hotter were offset by enhanced cooling at high densities that led to enhanced small-scale fragmentation.  On the other hand, \citet{Guszejnov_etal2022} found that subjecting solar-metallicity clouds to a stronger ISRF increased the characteristic stellar mass, which is in qualitative agreement with the results we present here.

\subsection{Extrapolation of the results to different redshifts}
\label{sec:extrapolate}

Fundamentally, the numerical results from this paper and \cite{Bate2019} point to an IMF that is relatively independent of metallicity (ranging from $z=0.01-3~{\rm Z}_\odot$) for present-day star formation ($z=0$), to one where the characteristic stellar mass increases with metallicity at high redshift (assuming all other parameters remain the same).  The main cause of this different qualitative behaviour is that for present-day star formation increasing the metallicity results in higher extinction of the ISRF and colder temperatures deep within star-forming clumps.  By contrast, at redshift $z=5$ the CMBR component of the ISRF that is able to penetrate into the densest regions of molecular cloud cores is hotter.  There is essentially no difference in the range of temperatures of high-density gas ($n_{\rm H} \gsim 10^8$~cm$^{-3}$) at low metallicity ($Z=0.01~{\rm Z}_\odot$) between the $z=0$ and $z=5$ calculations.  In the $z=0$ calculations the vast majority of the gas is much warmer than the temperature of the CMBR at $z=5$ ($T_{\rm CMBR}=(2.73~{\rm K}) (1+z)=16$~K), ranging between $T\approx 10-200$~K for densities $n_{\rm H} \gsim 10^8-10^{11}$~cm$^{-3}$.  Therefore, the hotter CMBR at $z=5$ has little effect and so the low-metallicity IMFs at $z=0$ and $z=5$ are almost identical.  However, at high metallicity (e.g.\ $Z={\rm Z}_\odot$), for $z=0$ a substantial fraction of the high-density gas has temperatures below 15~K and temperatures drop as low as $T\approx 6$~K.  Therefore, when otherwise identical calculations are carried out using the hotter CMBR at $z=5$ high-density gas is kept substantially hotter and fragmentation is reduced.  This leads to fewer protostars being formed and higher characteristic stellar masses because the protostars that do form are able to accrete more mass.

Based on these results, we can interpolate to determine how the IMF is likely to vary with metallicity and redshift between $z=0$ and $z=5$, and extrapolate for $z> 5$.  Since the CMBR temperature scales as $T_{\rm CMBR}=(2.73~{\rm K}) (1+z)$ we expect the IMF to be essentially identical to the $z=0$ cases up to $z=1$, since the CMBR temperature is lower than the minimum gas temperatures found in the $z=0$ calculations ($T_{\min}\approx 5$~K even for $Z=3~{\rm Z}_\odot$).  There would probably also be no significant change for $z=2$ either, because although a small amount of gas may otherwise be colder than $T_{\rm CMBR}$ it would be a negligible amount.  We base this on the fact that the minimum gas temperature for high-density gas is raised by about a factor of two between $z=0$ and $z=5$ for $Z=0.1~{\rm Z}_\odot$ but this results in a very small increase in the characteristic stellar mass (Table \ref{table1}).  For $z=3$, we would expect a very slight increase in the characteristic stellar mass (i.e.\ from $\approx 0.15$ to $\approx 0.20$~M$_\odot$ in terms of the mean of the logarithm of the stellar masses) for $Z\gsim 3~ {\rm Z}_\odot$, similar to that which is seen between $Z=0.01~{\rm Z}_\odot$ and $Z=0.1~{\rm Z}_\odot$ at $z=5$ because the minimum high-density gas temperature would be raised from $T\approx 5$~K to $T\approx 11$~K.  By redshift $z=4$ this magnitude of shift in the characteristic stellar mass should be apparent for solar metallicity, while for $Z=3~{\rm Z}_\odot$ the characteristic mass is likely to have doubled to $\approx 0.3$~M$_\odot$).

Beyond $z=5$ we would expect the characteristic stellar masses of the IMFs for all metallicities $Z> 0.01~{\rm Z}_\odot$ to increase with redshift.  As with the $z=5$ calculations, the greatest effect would initially be found in the highest metallicity clouds.  However, in the limit that $T_{\rm CMBR}$ becomes very large, say $\gsim 50$~K which is roughly the median temperature of the high density ($n_{\rm H} \gsim 10^8-10^{11}$~cm$^{-3}$) gas in all the of the calculations (Fig.~\ref{fig:phase}), then all the IMFs are likely become similar (and all with characteristic stellar masses significantly larger than the Galactic value).  Since the Jeans mass scales $\propto T^{3/2}\rho^{-1/2}$ it would be reasonable to expect that the characteristic mass (for clouds similar to those modelled here) would increase $\propto (1+z)^{3/2}$ in the limit of high redshift.

    Finally, a few remarks of caution.  First, all the calculations discussed here are performed in the regime where dust cooling dominates the cooling of high-density molecular gas.  We have intentionally limited our studies to metallicities $Z\ge 0.01~{\rm Z}_\odot$ for this reason.  In the limit of no metals, i.e., Population III stars, gas cooling is very different \citep[e.g.][]{BroCopLar1999,AbeBryNor2000}.  To model the very-low-metallicity regime (e.g. $Z=10^{-4}-10^{-2}~{\rm Z}_\odot$) accurately probably requires both metal line cooling of the gas and dust cooling to be modelled accurately \citep[e.g.][]{ShaKru2022}.  Second, to date we have performed such calculations using only a single type of initial condition -- relatively dense initial clouds ($n_{\rm H} \approx 6 \times 10^4$~cm$^{-3}$) with decaying turbulence and an initial root-mean-square Mach number of ${\cal M}=11.2$ at 15~K.  \cite{JonBat2018a} studied how the IMF may vary with cloud density for present-day star formation and found that the characteristic stellar mass varied relatively weakly with cloud density $\propto \rho^{-0.2}$.  However, other effects may come into play that we do not include, such as the effects of feedback from protostellar outfows \citep*[e.g.][]{MatFed2021,TanKruFed2022}.  All in all, there is still a great deal of work to be done in mapping out how the IMF may vary over a wide range of initial conditions.

\subsection{Observational evidence for a varying IMF}

\subsubsection{Galactic stellar populations}
\label{sec:galactic}

Although there is little direct evidence for variation of the IMF \citep*{BasCovMey2010}, there are a number of observations that provide indirect evidence in favour of variable IMFs.
\citet{Larson1998} lists several points including:  no metal-free stars have ever been found (implying that few, if any, of the first stars had masses $M\lsim 0.8~{\rm M}_\odot$); the `G-dwarf problem' in which the numbers of metal-poor stars in the solar neighbourhood are deficient compared to the predictions of simple chemical models \citep{vandenBergh1962,Schmidt1963}; the large total mass of heavy elements in large clusters of galaxies \citep{ElbArnVan1995,LoeMus1996,Gibson1997,GibMat1997};  and the observed increases of the metallicities (\citealt*{Faber1973,WorFabGon1992}; \citealt{Vazdekis_etal1996,Vazdekis_etal1997, Trager_etal2000,Gallazzi_etal2006}; \citealt*{GraFabSch2009a,GraFabSch2009b}; \citealt{Kuntschner_etal2010,McDermid_etal2015}) and mass-to-light ratios (\citealt*{GuzLucBow1993, JorFRaKja1996}; \citealt{Burstein_etal1997, Thomas_etal2011, Cappellari_etal2012, Cappellari_etal2013}; \citealt*{DutMenSim2012};,\citealt{Wegner_etal2012, Dutton_etal2013}; \citealt*{TorRomNap2013}; \citealt{McDermid_etal2014}; \citealt*{DavMcD2017}) of the central regions of early-type galaxies with increasing central velocity dispersion (i.e.\ mass).  Recently, gravitational lensing has also been used to measure the total projected mass within the Einstein radius to obtain mass-to-light ratios (\citealt*{FerSahWil2005, FerSahBur2008}; \citealt{Auger_etal2010, Ferreras_etal2010, Treu_etal2010, Barnabe_etal2011, Spiniello_etal2015, Newman_etal2017}).  Although the central regions of early-type galaxies are observed to have metallicities and mass-to-light ratios that increase with the central mass, observations of radial gradients that the outskirts of such galaxies are generally composed of metal-poor stars \citep[e.g.][]{Greene_etal2015}.

Historically, these observations and, in particular the high values of [$\alpha$/Fe] observed in massive early-type galaxies, have been used argue for a top-heavy IMF \citep[e.g.][]{Larson1998}.  \citet{Gunawardhana_etal2011} and \citet{Weidner_etal2013} have proposed that the IMF becomes more top-heavy with increasing star-formation rate.  However, \cite{Larson1998} argued instead that many of these observations can potentially be explained by IMFs that have the same (Salpeter-type) slope for high-mass stars, but a varying (characteristic) mass below which the IMF flattens or turns over.  For example, mass-to-light ratios of old stellar populations (e.g.\ those in elliptical galaxies) depend sensitively on the behaviour of the IMF in the vicinity of a solar-mass, since stars above $\approx 0.8$~M$_\odot$ have evolved into dark stellar remnants, while the light is generated by stars below this mass.  In general, mass-to-light ratios can be increased either by an excess of very low-mass stars (i.e.\ bottom-heavy), or an excess of stellar remnants (i.e.\ top heavy).

In principle, spectral analysis of stellar populations can be used to constrain directly the fraction of low-mass stars in early-type galaxies (\citealt*{Spinrad1962, Cohen1978, FabFre1980, CarVisPic1986, HarCou1988, DelHar1992}; \citealt{Cenarro_etal2003, Falcon_etal2003}; \citealt*{vanCon2010, vanCon2011, Convan2012a, Convan2012b, SmiLucCar2012}; \citealt{Spiniello_etal2012, Spiniello_etal2014};  \citealt{Ferreras_etal2013,  Ferreras_etal2015}; \citealt*{FerLaBVaz2015}; \citealt{LaBarbera_etal2013, LaBarbera_etal2017, MartinNavarro_etal2015, vanDokkum_etal2017}). 
Recent spectra analysis of the centres of early-type galaxies show a substantial population of low-mass stars \citep[e.g.][]{vanCon2010} and point to a bottom-heavy stellar population, although not in all early-type galaxies (\citealt*{SmiLuc2013, SmiLucCon2015}; \citealt{Leier_etal2016}).  \citet{Rosani_etal2018} find that need for a bottom-heavy IMF does not depend on the environment of the early-type galaxy (i.e.\ galaxy hierarchy or host halo mass).  To produce both the high metallicity and enhanced [$\alpha$/Fe] of early-type galaxies and a bottom-heavy IMF, a time-dependent IMF may be required that changes from top- to bottom-heavy during the early formation process \citep{Vazdekis_etal1997, Weidner_etal2013, Ferreras_etal2015}.

Recently, \citet*{GusHopGra2019} examined whether the near universal Milky Way IMF could be reconciled with proposed extragalactic IMF variations.  They found that it was very difficult to match the proposed extragalactic IMF variations without simultaneously predicting too much variation of the IMF in extreme Galactic environments.

Using the calculations presented in this paper, we cannot comment on variations in the slope of the IMF at the high-mass end since the calculations are only large enough to produce low- and intermediate-mass stars.  However, we can use them to predict the dependence of the characteristic stellar mass on both metallicity and redshift (Section \ref{sec:extrapolate}).  Unfortunately, it is not immediately clear how the predictions of this paper can be used to explain the above observations.  A characteristic stellar mass that increases for decreasing redshift and increasing metallicity would fit well with the picture presented by \citet{Larson1998} of a deficit of low-mass stars being able to explain many observations.  However, it does not appear to fit well with the recent spectral analysis results that point to an abundance of low-mass stars and a bottom-heavy IMF.  It may be that several different processes are contributing to the unusual stellar populations in the centres of early-type galaxies (e.g. high levels of turbulence or intense radiation fields from starbursts), rather than only changes due to the varying CMBR.  For example, the recent calculations of \citet{TanKruFed2022} show that at very high molecular cloud densities, feedback from protostellar outflows may result in an IMF with a lower characteristic stellar mass (and a narrower overall mass range) than the Galactic IMF.  Alternately, the results obtained in this paper do provide a mechanism for metal-rich gas to produce time-dependent IMFs that change from top-heavy to a standard IMF as redshift decreases \citep[similar to that postulated by][]{Vazdekis_etal1997, Weidner_etal2013, Ferreras_etal2015}.

\subsubsection{Globular clusters}

Globular clusters are much more simple objects than the central regions of galaxies.  They (largely) are thought to have formed from single star formation events, and they are not thought to contain significant dark matter.  When interpreting their stellar mass functions care does needs to be taken with the effects of dynamical evolution -- the lowest mass stars are lost preferentially.  But if this can be taken into account, they can in principle be used to study the form of the low-mass ($M\lsim 0.8$~M$_\odot$) IMF at high redshift.

In particular, a study of old globular clusters in M31 has found that more metal-rich globular clusters become increasingly bottom-light (i.e.\ have a deficit of low-mass stars).  \citet*{StrCalSet2011} studied more than a hundred globular clusters in M31, the vast majority of which were very old (i.e.\ formed at high redshift).  They divided their sample into several metallicity bins.  Their most metal-rich sample had metallicities [Fe/H]$~>-0.4$.  Crucially, \citet{StrCalSet2011} were able to detect that those clusters that were thought to be more dynamically-evolved were more bottom-light as expected due to dynamical evolution.  However, due to their large sample, they were also able to show that even accounting for the effects of dynamical evolution, globular clusters that were more metal-rich were increasingly bottom-light. 
There are questions over this result, however. \citet{ShaGie2015} claim that the masses of the M31 globular clusters were systematically underestimated and that the bias is more important at high metallicities.

Nevertheless, four of the M31 metal-rich globular clusters were also studied using the spectral analysis method of \citet{Convan2012a} by \citet{Convan2012b} to determine the distributions of low-mass stars.  In agreement with  \citet{StrCalSet2011} they found that they had a significant deficit of low-mass stars relative to a Galactic Kroupa IMF.

This observed trend of the stellar populations of more metal-rich globular clusters being more bottom-light is the opposite of the trends seen in spectral studies of the centres of early-type galaxies.  However, it is exactly the sort of trend that is expected from the numerical results presented in this paper.  Assuming that the old globular clusters were formed at redshifts of $z \gsim 5$, the results presented here can be used to predict that the characteristic stellar masses of old globular clusters should indeed have been higher for greater metallicity.  Similarly, the most metal-rich old globular clusters should have been born with deficits of low-mass stars compared to the present-day Galactic IMF.

\section{Conclusions}
\label{conclusions}

We have presented results from three radiation hydrodynamical simulations of star cluster formation that are subjected to the cosmic microwave background radiation at redshift $z=5$.  The three calculations have identical initial conditions except for their metallicity which ranges from 1/100 to 1 times the solar value.  The calculations resolve the opacity limit for fragmentation, protoplanetary discs (radii $\gsim 1$~au), and multiple stellar systems.  The calculations treat gas and dust temperatures separately and include a thermochemical model of the diffuse ISM that is important for setting the temperatures at low densities.  We compare the results presented in this paper with the results of present-day ($z=0$) star cluster formation calculations already published by \citet{Bate2019}.

We draw the following conclusions:
\begin{enumerate}
\item At redshift $z=5$ we find that the stellar mass functions produced by the calculations are become increasingly bottom-light with increasing metallicity.  In other words, the characteristic (median) stellar mass increases with increasing metallicity.  This behaviour is qualitatively different to that found for present-day ($z=0$) star formation in which the stellar mass function is found to be independent of metallicity (in the range $Z=0.01-3~{\rm Z}_\odot$).  At redshift $z=5$, the distribution of stellar masses at low metallicity ($Z=0.01~{\rm Z}_\odot$) is similar to the parametrisation of the observed Galactic IMF given by \citet{Chabrier2005}.  But at $z=5$ with solar metallicity the characteristic stellar mass is approximately 3 times higher.

\item Stellar multiplicity strongly increases with primary mass, similar to observed Galactic stellar systems.  But whereas at $z=0$ the multiplicity for a given primary mass is found to be similar for all metallicities ($Z=0.01-3~{\rm Z}_\odot$), at $z=5$ the multiplicity of M-dwarfs is found to decrease with increasing metallicity, such that their multiplicity is a factor of two lower at $Z={\rm Z}_\odot$ compared to $Z=0.01~{\rm Z}_\odot$.

\item At both $z=0$ and $z=5$ there is an anti-correlation between metallicity and the frequencies of close (semi-major axes $a<10$~au) bound protostellar pairs with primary masses $M_1=0.1-1.5~{\rm M}_\odot$.  However, the trend is not as strong at $z=5$ as at $z=0$, and the frequencies of close pairs are lower at $z=5$ than at $z=0$.

\item We also find that close (semi-major axes $a<10$~au) bound protostellar pairs have greater fractions of unequal-mass systems with lower metallicity at both $z=0$ and $z=5$.

\item The above differences between star formation at $z=5$ and $z=0$ are a result of metal-rich gas being unable to cool to as lower temperatures at $z=5$ as at $z=0$ due to the hotter cosmic microwave background radiation.  This inhibits fragmentation at high densities and high metallicity, meaning that the protostars that do form tend to accrete more gas (i.e.\ they have a greater characteristic stellar mass) and they have lower companion frequencies.

\item Comparing our results to observational evidence for a varying IMF, we find that the observed trend of old globular clusters in M31 being more bottom-light with increasing metallicity \citep{StrCalSet2011, Convan2012b} is consistent with the dependencies of the IMF on redshift and metallicity that we find from the numerical simulations.  The results may also help to explain some other observations that point to the need for varying IMFs, although they do not provide an explanation for bottom-heavy IMFs that are apparently observed in the centres of many massive elliptical galaxies.
\end{enumerate}

The calculations discussed in this paper begin with idealised initial conditions (i.e. uniform-density, spherical clouds initialised with a particular `turbulent' velocity field).  This is done to allow careful comparison between calculations to determine the effects that varying the CMBR and/or metallicity have. More realistic initial conditions (e.g. self-consistent turbulence and different cloud structures, such as turbulence driven by external supernovae; \citealt{Seifried_etal2018}) or different initial clouds (e.g. mean cloud densities; \citealt{JonBat2018a, TanKruFed2022}) may lead to some what different stellar properties.  Similarly, these calculations do not include magnetic fields or protostellar outflows.   Nevertheless, we expect the general trends for how stellar properties depend on redshift and metallicity that have been identified in this paper to be similar even if more complex initial conditions were employed or additional physical processes were to be included.

\section*{Acknowledgements}

MRB thanks Charlie Conroy for reading and commenting on a draft version of the manuscript. 

This work was supported by the European Research Council under the European Commission's Seventh Framework Programme (FP7/2007-2013 Grant Agreement No. 339248).  The calculations discussed in this paper were performed on the University of Exeter Supercomputer, Isca, and on the DiRAC Complexity system, operated by the University of Leicester IT Services, which forms part of the STFC DiRAC HPC Facility (www.dirac.ac.uk). The latter equipment is funded by BIS National E-Infrastructure capital grant ST/K000373/1 and STFC DiRAC Operations grant ST/K0003259/1. DiRAC is part of the National E-Infrastructure.  This research was supported in part by the National Science Foundation under Grant No. NSF PHY-1748958.  Some of the figures were produced using SPLASH \citep{Price2007}, an SPH visualization tool publicly available at http://users.monash.edu.au/$\sim$dprice/splash.

\section*{Data availability}

Data that can be used to produce Table \ref{tablemult} and Figs.~\ref{fig:IMF} to \ref{fig:CBq}, and to calculate many of the values in Table \ref{table1} are provided as Additional Supporting Information (see below).  
Input and output files from the three calculations that were performed for this paper are available from the University of Exeter's Open Research Exeter (ORE) repository \citep{Bate2022b_data}.  This dataset includes the initial conditions and input files for the three SPH calculations, the SPH dump files that were used to create Figs.~\ref{fig:DTZ001}--\ref{fig:chemistrysnaps}, and the output files that were used to produce Fig.~\ref{massnumber} and Tables \ref{table1}, \ref{tablestars}, and \ref{tablemultprop}. 

\section*{Supporting Information}

Additional Supporting Information may be found in the online version of this article:

{\bf Data files for protostars.} We provide text files of Tables \ref{tablestars} and \ref{tablemultprop} that give the properties of the protostars and multiple systems for each of the three calculations.  These files contain the data necessary to construct Figs.~\ref{fig:IMF} to \ref{fig:CBq}, and to produce Table \ref{tablemult}.  Their file names are of the format {\tt Table3\_Stars\_z5\_MetalX.txt} and  {\tt Table4\_Multiples\_z5\_MetalX.txt} where `X' gives the metallicity (`001' for 0.01, `01' for 0.1, or `1').  

{\bf Animations.} We provide animations of the evolution of the column density, and the gas, dust, and radiation temperatures for each of the three calculations (i.e.\ 12 animations).  Sink particles are represented by white circles.  These animations are provided to support the snapshots provided in Figs.~\ref{fig:DTZ001} to \ref{fig:DTZ1}.  Their file names are of the format {\tt Bate2022\_z5\_MetalX\_Y.txt} where `X' gives the metallicity and `Y' gives the variable (`Density', `Tgas', or `Tdust' or `Trad').

\bibliography{/Users/mbate/Tex/mbate}

\end{document}